\documentclass[twocolumn,showpacs,prc,superscriptaddress]{revtex4}

\usepackage{graphicx}
\usepackage{dcolumn}
\usepackage{bm}
\usepackage{tabularx}

\usepackage{mathrsfs}

\def\m@thcombine#1#2{%
  \setbox0=\hbox{$#1$}
  \setbox1=\hbox{$#2$}
  \ifdim\wd0>\wd1
    \setbox0=\hbox to\wd1{\hss\box0\hss}
  \else
    \setbox1=\hbox to\wd0{\hss\box1\hss}
  \fi
  \mathop{\vcenter{
    \offinterlineskip\box0\box1}}}
\def\lesim{\m@thcombine<\sim}
\def\gesim{\m@thcombine>\sim}

\def\vec#1{\mbox{\boldmath $#1$}}

\newcommand{\bra}{\langle}
\newcommand{\ket}{\rangle}

\newcommand{\hs}{\hspace{30pt}}

\newcommand{\vecr}{\boldsymbol r}

\begin{document}

\title{Continuum Hartree-Fock-Bogoliubov theory for 
weakly bound deformed nuclei using coordinate-space Green's function method}

\author{Hiroshi Oba}
\email{ooba@nt.sc.niigata-u.ac.jp}
\affiliation{Graduate School of Science and Technology, Niigata University, Niigata 950-2181, Japan}

\author{Masayuki Matsuo}%
\affiliation{Department of Physics, Faculty of Science, Niigata University, Niigata 950-2181, Japan}

\begin{abstract}
We formulate a new scheme of the Hartree-Fock-Bogoliubov mean-field theory applicable to weakly bound
and pair correlated 
deformed nuclei using the coordinate-space Green's function technique.  On the basis of a 
coupled-channel representation of the quasiparticle wave function expanded in terms
of the partial waves, we impose 
the correct boundary condition of the asymptotically out-going waves
on the continuum quasiparticle states.
We perform numerical analysis for $^{38}$Mg to illustrate properties of the
continuum quasiparticle states and the pair correlation 
in deformed nuclei near the neutron drip-line. 
\end{abstract}

\date{\today}

\begin{abstract}
We formulate a new scheme of the Hartree-Fock-Bogoliubov mean-field theory applicable to weakly bound
and pair correlated 
deformed nuclei using the coordinate-space Green's function technique.  On the basis of a 
coupled-channel representation of the quasiparticle wave function expanded in terms
of the partial waves, we impose 
the correct boundary condition of the asymptotically out-going waves
on the continuum quasiparticle states.
We perform numerical analysis for $^{38}$Mg to illustrate properties of the
continuum quasiparticle states and the pair correlation 
in deformed nuclei near the neutron drip-line. 
\end{abstract}

\maketitle

\section{Introduction}

The RI-beam facilities in the new generation will
enlarge significantly the experimentally accessible region in the nuclear 
chart, in particular in medium and heavy mass domains. An interesting area may be
the $10\lesim Z < 20$ and $N \gesim 20$ region, where the selfconsistent mean-field
theories predict that the shape deformation systematically occurs 
even if nuclei are close to the 
drip-line\cite{terasaki,lalazissis,stoitsov,stoitsov2,THO_table,Rodriguez,Gauss_HFB_2}.
It is the presence of weakly bound neutrons that makes these nuclei interesting, and furthermore
the possible shape deformation will bring about additional mechanism influencing 
the single-particle motion and the many-body correlations such as the
pairing and the collective excitations. 

A promising theoretical framework to describe this situation may be the
selfconsistent mean-field approach\cite{SC_mean-field,PVKC07,RingSchuck}. We consider more specifically the 
Hartree-Fock-Bogoliubov (HFB) method to describe the pair correlated and
deformed ground state\cite{SC_mean-field,RingSchuck}, 
and the quasiparticle random phase approximation (QRPA)
to describe excitation modes built on the ground state\cite{PVKC07,RingSchuck}. 
Note here that, since the nucleons
are bound only weakly and the threshold energy for the nucleon separation is low, 
one has to formulate the HFB and QRPA methods on the basis of correct description of 
the asymptotic forms of the wave functions 
of weakly bound and unbound continuum quasiparticle 
states\cite{1DHFB_BUL,1DHFB_DOBA,Green_sph_HFB,cQRPA_Matsuo,Khan}. 
The formalisms that fulfill this requirement, which we shall call
the continuum HFB\cite{Green_sph_HFB,Fayans,Grasso,Gamow_HFB,PTG_HFB} 
or the continuum QRPA\cite{cQRPA_Matsuo,Khan},
are limited mostly to spherical nuclei. Therefore we need 
to extend, as the first step, the continuum
 HFB to deformed nuclei, i.e., we need
to formulate the deformed continuum HFB theory. 
Hamamoto \cite{hamamoto_def_HFB_1,hamamoto_def_HFB_2} has analyzed
in detail the quasiparticle motion 
in deformed 
Woods-Saxon potential by solving the HFB equation in the coupled-channel representation
imposing the boundary condition of the 
correct asymptotics.
Recently Stoitsov et al. \cite{PTG_HFB} have introduced 
a formulation of the deformed continuum HFB  
utilizing the P\"{o}schel-Teller-Ginochio basis. 
In the present paper, we intend to give another new formulation of
the deformed continuum HFB method by extending the Green's function approach\cite{Green_sph_HFB}.
Our eventual target is not just to formulate the deformed continuum HFB, but also 
to formulate a continuum QRPA for deformed nuclei.  Although the latter is not 
pursued here, the deformed continuum QRPA can be easily formulated once
the quasiparticle Green's function for deformed nuclei is constructed
in the coordinate representation\cite{cQRPA_Matsuo}.

The kernel of the present formulation is the quasiparticle 
Green's function (called also the HFB Green's function) 
which satisfies the correct boundary conditions of the continuum
quasiparticle states. This can be achieved by utilizing the
coupled-channel representation\cite{hamamoto_def_HFB_1,hamamoto_def_HFB_2} 
of the quasiparticle Schr\"{o}dinger equation (the HFB equation)
based on the partial wave expansion.
The exact form of the HFB Green's function is known for spherical 
nuclei\cite{Green_sph_HFB} where the channels 
- the partial waves - decouple, but  
what we need is the one in deformed nuclei for which there exists the coupling among the partial
waves. We mention here that the Green's function for deformed potentials has been
utilized in  describing other physical systems, e.g. the electronic response in 
molecules\cite{Levine-Soven,Levine}
and electrons in matter
scattering on deformed ion potentials\cite{Foulis},
for which the coupled-channel representation based
on the partial wave expansion is also employed. 
Indeed 
the exact form of the Green's function in the general coupled-channel system
\cite{Foulis} can be extended to our problem, i.e. to describe 
the quasiparticle wave functions in pair correlated deformed nuclei. 
In the HFB theory, one also needs to calculate 
the density and the pair density, or the generalized density matrix in general, by
summing up the wave functions of all the quasiparticle states including those in the
continuum. The Green's function formalism\cite{Green_sph_HFB} can be utilized also
to efficiently perform
this summation. In this way we obtain a complete scheme to calculate
the HFB ground state and the single-particle properties influenced by the pair correlation.
We explain the details of this formalism in Section II.

As a demonstration of the deformed continuum HFB method, we perform in section III numerical analysis
for a neutron-rich nucleus $^{38}$Mg, which is situated near the neutron
drip-line and is predicted to be prolately deformed in many selfconsistent mean-field
calculations\cite{terasaki,lalazissis,stoitsov,stoitsov2,THO_table,Rodriguez,Gauss_HFB_2}.  
 The purpose of this analysis is two fold. The first is to 
investigate how the neutron pair correlation changes as the binding of neutrons becomes
weaker and stronger.  Secondly we would like to reveal peculiar
properties of the single-particle motion in the pair correlated 
deformed nuclei near the neutron drip-line, where the coupling of the
quasiparticle states to the continuum orbits may introduce  new features in
the single-particle motion.
We shall compare our results with those of Hamamoto \cite{hamamoto_def_HFB_1,hamamoto_def_HFB_2}, 
who performed a pioneering
analysis of the continuum quasiparticle states in a formalism satisfying the correct asymptotic forms,
but not on the basis of the selfconsistent treatment of the pair correlation.
Section IV is devoted to the conclusions.

\section{Deformed continuum HFB theory in the Green's function formalism}

\subsection{Coordinate-space HFB equation}

The Bogoliubov's quasiparticle plays a central role
in the HFB theory\cite{RingSchuck}. The wave function
of the quasiparticle state has two components, and is
written in the coordinate-space representation as
\begin{equation}
\phi(\vecr \sigma)=
\left (
\begin{array}{c}
\phi^{(1)}(\boldsymbol{r} \sigma) \\
\phi^{(2)}(\boldsymbol{r} \sigma)
\end{array}
\right ).
\end{equation}
It obeys the HFB equation
\begin{equation}
\left ( 
\begin{array}{cc}
 h-\lambda & \ \tilde{h} \\ 
 \tilde{h} & -h+\lambda 
\end{array}
\right )
\left (
\begin{array}{c}
\phi^{(1)}(\boldsymbol{r} \sigma, E) \\
\phi^{(2)}(\boldsymbol{r} \sigma, E)
\end{array}
\right )
= 
E 
\left (
\begin{array}{c}
\phi^{(1)}(\boldsymbol{r} \sigma, E) \\
\phi^{(2)}(\boldsymbol{r} \sigma, E)
\end{array}
\right ),
\label{hfbeq0}
\end{equation}
where $E$ is the quasiparticle excitation energy, and 
$\lambda$ is the chemical potential or the Fermi energy which should be
determined to constrain the expectation value of the 
nucleon number. We omit the isospin index for simplicity of
notation. Here and hereafter, we assume the 
time reversal invariance of the HFB ground state $ | \Psi \ket $,
the associated Hartree-Fock Hamiltonian $h$ and
the pair Hamiltonian $\tilde{h}$. We also make a simplification
that the 
Hartree-Fock potential and the pair Hamiltonian are
local ones:
\begin{equation}
h
= 
-\frac{\hbar^2}{2m}\nabla^2
+V(\boldsymbol{r} \boldsymbol{s}), \ \ \ \ 
\tilde{h}=\Delta(\boldsymbol{r}),
\end{equation}
as  is realized in the cases of some Skyrme 
effective interactions, e.g. SkP\cite{1DHFB_DOBA} with the
effective mass $m^* =1$, or of the phenomenological
Woods-Saxon potential. The pair Hamiltonian becomes a local
pair potential $\Delta(\boldsymbol{r})$ when the contact
force is adopted for the pairing interaction. 
The position dependent effective mass $m^*(\vecr)$ 
is often encountered in the Skyrme Hartree-Fock(-Bogoliubov) theories.
An extension to this
case is straightforward,
but we do not deal with here. 

We expand the quasiparticle wave function with respect to
the partial waves specified by the angular quantum numbers
$jlm$, abbreviated by $L$ hereafter:
\begin{equation}
\left (
\begin{array}{c}
\phi^{(1)}(\boldsymbol{r} \sigma, E) \\
\phi^{(2)}(\boldsymbol{r} \sigma, E)
\end{array}
\right )
=
\sum_{L}
\left (
\begin{array}{c}
\phi^{(1)}_{L}(r, E) Y_{L}(\hat{r}\sigma) \\
\phi^{(2)}_{L}(r, E) Y_{L}(\hat{r}\sigma) \\
\end{array}
\right ),
\end{equation}
where $Y_{L}(\hat{r}\sigma)$ is the spin spherical harmonics.
The HFB equation is then transformed to a 
coupled-channel form\cite{hamamoto_def_HFB_1,hamamoto_def_HFB_2} for the
radial wave functions $\phi^{(i)}_{L}(r, E)$:
\begin{widetext}
\begin{eqnarray}
\left (
-\frac{\hbar^2}{2m}\frac{\partial^2}{{\partial r}^2 }
-\frac{\hbar^2}{2m}\frac{2}{r}\frac{\partial}{\partial r}
+\frac{\hbar^2}{2m}\frac{\ell(\ell+1)}{r^2}
-\lambda -E
\right) \phi^{(1)}_{L}(r,E)
&+&\sum_{L'} \left (
u^{(0)}_{LL'}(r) + u^{(1)}_{LL'}(r)\frac{\partial}{\partial r}
\right ) \phi^{(1)}_{L'}(r,E)
\nonumber \\
&+&\sum_{L'}
\Delta_{LL'}(r) \phi^{(2)}_{L'}(r,E)
= 0, \label{coupled_channel_eq_1}
\\
- \left (
-\frac{\hbar^2}{2m}\frac{\partial^2}{{\partial r}^2 }
-\frac{\hbar^2}{2m}\frac{2}{r}\frac{\partial}{\partial r}
+\frac{\hbar^2}{2m}\frac{\ell(\ell+1)}{r^2}
-\lambda +E
\right) \phi^{(2)}_{L}(r,E)
&-&\sum_{L'} \left (
u^{(0)}_{LL'}(r) + u^{(1)}_{LL'}(r)\frac{\partial}{\partial r}
\right ) \phi^{(2)}_{L'}(r,E)
\nonumber \\
&+&\sum_{L'}
\Delta_{LL'}(r) \phi^{(1)}_{L'}(r,E)
= 0. \label{coupled_channel_eq_2}
\end{eqnarray} 
\end{widetext}
Here the ``channels" are
labeled by the quantum number $jlm(=L)$ and the index $i=1,2$
specifying the upper and lower components of the 
quasiparticle wave function.  The coupling among
the channels is governed by
$u^{(0)}_{LL'}(r) \,,\,
u^{(1)}_{LL'}(r)$ and 
$\Delta_{LL'}(r),$ which are defined by
\begin{equation}
u^{(0)}_{LL'}(r)+
u^{(1)}_{LL'}(r)\frac{\partial}{\partial r}
=
\int d\hat{r}\sum_{\sigma\sigma'}Y_{L}^{*}(\hat{r}\sigma') V(\boldsymbol{r} \boldsymbol{s}) 
Y_{L'}(\hat{r}\sigma) ,
\label{dot_Vr}
\end{equation}
\begin{equation}
\Delta_{LL'}(r)
=
\int d\hat{r}\sum_{\sigma} Y_{L}^{*}(\hat{r}\sigma) \Delta(\boldsymbol{r}) Y_{L'}(\hat{r}\sigma) .
\label{dot_deltar}
\end{equation}
Note that the coupled-channel representation is often employed to
describe scattering states in the non-spherical 
potential problems\cite{Levine-Soven,Levine,Foulis,Hagino,hamamoto_def_WS}.
We have to truncate the partial wave expansion in practical calculations, and we
denote $N$ to represent the number of partial waves to be included.

Using the two component radial wave function
\begin{equation}
\phi_L(r,E) = 
\left (
\begin{array}{c}
\phi^{(1)}_{L}(r,E)  \\
\phi^{(2)}_{L}(r,E)  \\
\end{array}
\right ),
\end{equation}
the coupled-channel equation can be written as
\begin{widetext}
\begin{equation}
\frac{\partial^2}{{\partial r}^2 }\phi_L(r,E)
+\frac{2}{r}\frac{\partial}{\partial r}\phi_L(r,E)
+\sum_{L'}v_{LL'}^{(1)}(r)\frac{\partial}{\partial r}\phi_{L'}(r,E)
+\sum_{L'}v_{LL'}^{(0)}(r,E)\phi_{L'}(r,E)
= 0,
\end{equation}
where
we introduced $2 \times 2$ matrices
\begin{equation}
v_{LL'}^{(0)}(r,E) = \left ( 
\begin{array}{cc}
-\frac{\ell(\ell+1)}{r^2} - \alpha u_{LL'}^{(0)}(r) +\alpha \lambda +\alpha E &
-\alpha \Delta_{LL'}(r) \\
\alpha \Delta_{LL'}(r) &
-\frac{\ell(\ell+1)}{r^2} - \alpha u_{LL'}^{(0)}(r) +\alpha \lambda -\alpha E 
\end{array}
\right ),
\end{equation}
\end{widetext}
\begin{equation}
v_{LL'}^{(1)}(r) =  \left (
\begin{array}{cc}
-\alpha u_{LL'}^{(1)}(r) & 0 \\
 0 & -\alpha u_{LL'}^{(1)}(r)
\end{array}
\right ),
\end{equation}
and a constant $\alpha = 2m/\hbar^2$.
It is also possible to write this equation in a grand matrix form
\begin{equation}
\left (
\frac{\partial^2}{{\partial r}^2 }
+\frac{2}{r}\frac{\partial}{\partial r}
+\mathrm{v}^{(1)}(r)\frac{\partial}{\partial r}
+\mathrm{v}^{(0)}(r,E)
\right )
\phi(r,E)
= 0,
 \label{hfb_matrix}
\end{equation}
where the radial wave functions
with different $L$'s are combined to form a $2N$ dimensional vector
\begin{equation}
\phi(r,E)=
\left(
\begin{array}{c}
\phi_{L_1}(r,E) \\
\phi_{L_2}(r,E) \\
\vdots \\
\phi_{L_N}(r,E)
\end{array}
\right ),
\end{equation}
and $2N \times 2N$ matrices $\mathrm{v}^{(0)}$ and $\mathrm{v}^{(1)}$ are defined by
\begin{equation}
[\mathrm{v}^{(n)}]_{LL'} = v_{LL'}^{(n)}. \hs (n=1,2)
\end{equation}

\subsection{Boundary conditions for the quasiparticle wave function}

The quasiparticle states with the energy higher than the nucleon
separation energy form a continuum spectrum\cite{1DHFB_BUL,1DHFB_DOBA,Green_sph_HFB}.
They are the states with $|E|>|\lambda|$ while the discrete quasiparticle states lie in the
energy range $|E|<|\lambda|$. In order to describe the continuum quasiparticle
states, it is an essential condition that
the
quasiparticle wave functions  must have
correct asymptotic forms
at far outside $r \rightarrow \infty$  where the potentials
vanish.

We therefore impose the boundary condition at $r \rightarrow \infty$ that
the radial wave functions should be connected to the following asymptotic form.
There are $2N$ independent solutions where the upper or lower component
of a given partial wave $L'$ is dominant:
\begin{equation}
\phi_{L}(r,E)  
\stackrel{r \to \infty }{\longrightarrow}
\left (
\begin{array}{c}
\frac{H^{+}_{l}(k_{+},r)}{r} \delta_{LL'} \\
0 
\end{array}
\right ),
\  \
\left (
\begin{array}{c} 
0 \\
\frac{H^{+}_{l}(k_{-},r)}{r} \delta_{LL'}
\end{array}
\right ).
\label{asympt}
\end{equation}
Here 
\begin{equation}
H^{+}_{l}(k,r) = \left \{
\begin{array}{cc}
j_{l}(k,r) + i n_{l}(k,r) &  (\mbox{for neutron }) \\
F_{l,Z}(k,r) + i G_{l,Z}(k,r) &  (\mbox{for proton }) 
\end{array}
\right .  
\end{equation}
is the Hankel/Coulomb function. The wave numbers in Eq.~(\ref{asympt})
are given by
$k_{\pm}(E)=\sqrt{2m(\lambda \pm E)} / \hbar$
and their branch cuts are chosen so that ${\rm Im}k_{\pm}(E) >0$
is satisfied\cite{Green_sph_HFB}. This boundary condition is equivalent 
to imposing that the wave functions 
are connected to the asymptotic out-going or exponentially decaying waves.
Combining these solutions in columns, we introduce a $2\times 2$ form
$\varphi^{(out)}_{LL'}(r,E)$ satisfying
\begin{equation}
\varphi_{LL'}^{(out)}(r,E)  
\stackrel{r \to \infty }{\longrightarrow}
\left (
\begin{array}{cc}
\frac{H^{+}_{l}(k_{+},r)}{r} \delta_{LL'} &
0 \\
0 &
\frac{H^{+}_{l}(k_{-},r)}{r} \delta_{LL'}
\end{array}
\right ), \label{BC_asympt}
\end{equation}
and similarly a $2N \times 2N$ matrix form $\Phi^{(out)}(r,E)$ defined by
\begin{equation}
[\Phi^{(out)}(r,E)]_{LL'} = \varphi^{(out)}_{LL'}(r,E).
\end{equation}

At the center of
the nucleus, 
we consider the radial wave functions which are regular at 
$r=0$:
\begin{equation}
\varphi_{LL'}^{(in)}(r,E) 
\stackrel{r \to 0 }{\longrightarrow}
\left (
\begin{array}{cc}
r^{l} \delta_{LL'} &
0 \\
0 &
r^{l} \delta_{LL'}
\end{array}
\right ). \label{BC_origin}
\end{equation}
We denote it also in the $2N \times 2N$ matrix form as
\begin{equation}
[\Phi^{(in)}(r,E)]_{LL'} = \varphi^{(in)}_{LL'}(r,E) .
\end{equation}

\subsection{HFB Green's function in the coupled-channel representation}

Now we introduce the Green's function 
 defined for the coordinate-space HFB
equation (\ref{hfbeq0}). It is expressed formally by
\begin{equation}
G(E) \equiv
\left [
E
-
\left ( 
\begin{array}{cc}
 h-\lambda & \ \tilde{h} \\ 
 \tilde{h} & -h+\lambda 
\end{array}
\right )
\right ]^{-1},
\end{equation}
and may be denoted  
\begin{equation}
G(\boldsymbol{r}\sigma,\boldsymbol{r}'\sigma',E) = \left (
\begin{array}{cc}
G^{(11)}(\boldsymbol{r}\sigma,\boldsymbol{r}'\sigma',E) &
G^{(12)}(\boldsymbol{r}\sigma,\boldsymbol{r}'\sigma',E) \\
G^{(21)}(\boldsymbol{r}\sigma,\boldsymbol{r}'\sigma',E) &
G^{(22)}(\boldsymbol{r}\sigma,\boldsymbol{r}'\sigma',E)
\end{array}
\right )
\end{equation}
in the coordinate-space
representation. Note that the HFB Green's function $G(E)$ has the $2\times 2$
matrix form, whose diagonal and off-diagonal components are often referred to as
the normal and abnormal Green's function, respectively.
Using the partial wave expansion it may be expanded as
\begin{equation}
G(\boldsymbol{r}\sigma,\boldsymbol{r}'\sigma',E) =
\sum_{LL'}Y_{L}(\hat{r}\sigma) g_{LL'}(r,r',E)Y_{L'}^{*}(\hat{r}'\sigma'),
\end{equation}
\begin{equation}
g_{LL'}(r,r',E) = \left (
\begin{array}{cc}
g^{(11)}_{LL'}(r,r',E) &
g^{(12)}_{LL'}(r,r',E) \\
g^{(21)}_{LL'}(r,r',E) &
g^{(22)}_{LL'}(r,r',E)
\end{array}
\right ).
\end{equation}
We can also introduce
the $2N \times 2N $  matrix form $\mathrm{g}(r,r',E)$ of $g_{LL'}(r,r',E)$:
\begin{equation}
[\mathrm{g}(r,r',E)]_{LL'} = g_{LL'}(r,r',E).
\end{equation}
It is easy to derive from the definition of the Green's function that
$\mathrm{g}(r,r',E)$ must satisfy the equation
\begin{eqnarray}
\left (
\frac{\partial^2}{{\partial r}^2 }
+\frac{2}{r}\frac{\partial}{\partial r}
+\mathrm{v}^{(1)}(r)\frac{\partial}{\partial r}
+\mathrm{v}^{(0)}(r,E) \right) \mathrm{g}(r,r',E)  \nonumber \\
= \frac{\delta(r-r')}{rr'}\alpha J,
 \label{hfbgf_matrix}
\end{eqnarray}
where $J$ is defined by
\begin{equation}
J= \left (
\begin{array}{cc}
1 & 0 \\
0 & -1
\end{array}
\right )\otimes I_N
\end{equation}
with $I_N$ being the $N$ dimensional unit matrix. 

We now seek the Green's function $\mathrm{g}(r,r',E)$ which
satisfies both Eq.(\ref{hfbgf_matrix}) and the boundary conditions, (\ref{BC_asympt}) 
and (\ref{BC_origin}), 
imposed at $r \rightarrow \infty$ and $r=0$, respectively. According to 
Ref.\cite{Foulis}, we write it in a form
\begin{eqnarray}
g_{LL'}(r,r',E) = \sum_{L''} \left (
C^{(in)}_{L'L''}(r',E)\varphi^{(in)}_{LL''}(r,E) \theta(r'-r) \right .
\nonumber \\
+ 
\left .
C^{(out)}_{L'L''}(r',E)\varphi^{(out)}_{LL''}(r,E) \theta(r-r')
\right ), \label{gf_cc}
\end{eqnarray}
or equivalently
\begin{eqnarray}
\mathrm{g}(r,r',E) = \Phi^{(in)}(r,E) {\mathrm{C}^{(in)}}^{T}(r',E) \theta(r'-r)
\nonumber\\
+ \Phi^{(out)}(r,E) {\mathrm{C}^{(out)}}^{T}(r',E) \theta(r-r') \label{gf_matrix},
\end{eqnarray}
where $\mathrm{C}^{(in)}(r,E)$ and $\mathrm{C}^{(out)}(r,E)$ are
matrix functions 
\begin{equation}
[\mathrm{C}^{(in/out)}(r,E)]_{LL'} = C^{(in/out)}_{LL'}(r,E).
\end{equation}
Substitution of Eq.(\ref{gf_matrix}) into Eq.(\ref{hfbgf_matrix}) leads to the equations for 
$\mathrm{C}^{(in)}(r,E)$ and $\mathrm{C}^{(out)}(r,E)$:
\begin{eqnarray}
\left (
\begin{array}{cc}
{\Phi^{(in)}}(r,E) & 
-{\Phi^{(out)}}(r,E) \\
-{\frac{d}{dr}\Phi^{(in)}}(r,E) &
{\frac{d}{dr}\Phi^{(out)}}(r,E)
\end{array}
\right )
\left (
\begin{array}{c}
{\mathrm{C}^{(in)}}^{T}(r,E) \\
{\mathrm{C}^{(out)}}^{T}(r,E)
\end{array}
\right )
\nonumber \\
= \frac{\alpha}{{r}^2}
\left (
\begin{array}{c}
0 \\
J
\end{array}
\right )
\label{hfbgf_matrix_7}
\end{eqnarray}
which corresponds to Eq.(23) of Ref.\cite{Foulis}. If
the Hamiltonian does not contain the coupling term with the first derivative,
i.e.  $\mathrm{v}^{(1)}(r)=0$, the solution has a simple form
\begin{eqnarray}
{\mathrm{C}^{(in)}}^T(r,E)&=&\alpha \left(\mathrm{M}^T\right)^{-1}
{\Phi^{(out)}}^T(r,E),
\nonumber \\
{\mathrm{C}^{(out)}}^T(r,E)&=&\alpha \mathrm{M}^{-1}
{\Phi^{(in)}}^T(r,E),
\end{eqnarray}
with the Wronskian matrix defined by
\begin{eqnarray}
\mathrm{M}(E)=r^2
\left[
{\Phi^{(in)}}^T(r,E) J \left( \frac{d}{dr} {\Phi^{(out)}}(r,E)\right)
\right .
\nonumber\\
\left .
-  
\left(\frac{d}{dr} {\Phi^{(in)}}^T(r,E)\right) J {\Phi^{(out)}}(r,E)
\right],
\end{eqnarray}
and consequently the Green's function is written as
\begin{eqnarray}
\mathrm{g}(r,r',E) &=&  
\alpha
\left(
\Phi^{(in)}(r,E) \left(\mathrm{M}^T\right)^{-1}{\Phi^{(out)}}^{T}(r',E)
\theta(r'-r)
\right .
\nonumber \\
 &+& 
\left . 
\Phi^{(out)}(r,E) \mathrm{M}^{-1}{\Phi^{(in)}}^{T}(r',E) 
\theta(r-r')
\right),
 \label{gf_matrix_2}
\end{eqnarray}
in parallel to Eq.~(36) of Ref.\cite{Foulis}.

In the following numerical application, however, we do not use
Eq.(\ref{gf_matrix_2}), but instead we solve directly Eq.(\ref{hfbgf_matrix_7}) to obtain
$\mathrm{C}^{(in)}(r,E)$ and $\mathrm{C}^{(out)}(r,E)$. We then  
use Eq.(\ref{gf_cc}) to
evaluate the HFB Green's function. This procedure is applicable to the case where
the coupling with the first derivative is present. We found also
that the procedure provides us better numerical accuracy than to use Eq.(\ref{gf_matrix_2})
even when there is no first derivative term.

\subsection{Generalized density matrix}

Key quantities in the HFB theory are the normal density matrix
\begin{equation}
\rho(\boldsymbol{r}\sigma,\boldsymbol{r}'\sigma')  
= \bra \Psi | \psi^\dagger(\boldsymbol{r}'\sigma')\psi(\boldsymbol{r}\sigma) | \Psi \ket,
\end{equation}
and the abnormal density (pair density) matrix
\begin{equation}
\tilde{\rho}(\boldsymbol{r}\sigma,\boldsymbol{r}'\sigma')
=
\bra \Psi | \psi(\boldsymbol{r}'\tilde{\sigma}')\psi(\boldsymbol{r}\sigma) | \Psi \ket,
\end{equation}
which can be combined as the generalized density matrix 
\begin{equation}
 R(\boldsymbol{r}\sigma,\boldsymbol{r}'\sigma') = 
 \left (
 \begin{array}{cc}
  \rho(\boldsymbol{r}\sigma,\boldsymbol{r}'\sigma') & \tilde{\rho}(\boldsymbol{r}\sigma,\boldsymbol{r}'\sigma') \\
  \tilde{\rho}^{*}(\boldsymbol{r}\tilde{\sigma},\boldsymbol{r}'\tilde{\sigma}')
   & \delta_{\boldsymbol{r}\boldsymbol{r}'}\delta_{\sigma \sigma'}-\rho^{*}(\boldsymbol{r}\tilde{\sigma},\boldsymbol{r}'\tilde{\sigma}')
 \end{array}
 \right ).
\end{equation}
Let $i$ be the index to specify the quasiparticle states, and $\phi_i(\vecr\sigma)$ be the eigen solutions
of the HFB equation (\ref{hfbeq0}). Then the generalized density matrix is
 obtained by summing up products of the
quasiparticle wave functions over all the quasiparticle states \cite{cQRPA_Matsuo}:
\begin{equation}
 R(\boldsymbol{r}\sigma,\boldsymbol{r}'\sigma') = 
 \sum_{i} \bar{\phi}_{\tilde{i}}(\boldsymbol{r}\sigma)\bar{\phi}_{\tilde{i}}^{\dag}(\boldsymbol{r}'\sigma')
\end{equation}
where $\bar{\phi}_{\tilde{i}}$ is a conjugate wave function of $\phi_i$, defined by
\[
\bar{\phi}_{\tilde{i}}(\boldsymbol{r}\sigma) \equiv
 \left ( 
 \begin{array}{c}
  -\phi_{i}^{(2)*}(\boldsymbol{r}\tilde{\sigma}) \\
  \phi_{i}^{(1)*}(\boldsymbol{r}\tilde{\sigma})
 \end{array}
\right) 
=
 \left ( 
 \begin{array}{cc}
  0 & -1 \\
  1 & 0
 \end{array}
\right) 
\phi_{\tilde{i}}(\boldsymbol{r}\sigma).
\]
Here $\tilde{\sigma}$ implies $\phi_i^{(i)}(\boldsymbol{r}\tilde{\sigma}) \equiv
(-2\sigma)\phi_i^{(i)}(\boldsymbol{r}-\sigma)$, and $\phi_{\tilde{i}}^{(i)}(\boldsymbol{r}\sigma)$
is the time-reversal  $\phi_{\tilde{i}}^{(i)}(\boldsymbol{r}\sigma)\equiv
T\phi_i^{(i)}(\vecr\sigma)=\phi_i^{(i)*}(\vecr\tilde{\sigma})$ of $\phi_i^{(i)}(\vecr\sigma)$.

We perform this summation by using a contour integral of the HFB Green's function
in the complex plane of the quasiparticle energy $E$ \cite{Green_sph_HFB}. Then the
generalized density matrix is given by 
\begin{equation}
 R(\boldsymbol{r}\sigma,\boldsymbol{r}'\sigma') 
= \frac{1}{2\pi i} \int_{C}
G(\boldsymbol{r}\sigma,\boldsymbol{r}'\sigma',E) dE .
\end{equation}
We use the
partial wave expansion also for the generalized density matrix:
\begin{equation}
R(\boldsymbol{r}\sigma,\boldsymbol{r}'\sigma')  
=\sum_{LL'} Y_{L}(\hat{r}\sigma) R_{LL'}(r,r')Y_{L'}^{*}(\hat{r}'\sigma'),
\end{equation}
\begin{equation}
R_{LL'}(r,r') = \frac{1}{2\pi i} \int_{C}
g_{LL'}(r,r',E) dE.
\end{equation}
The contour $C$ should enclose the negative energy side
of the real $E$ axis. The use of the contour
integral enables us to impose the proper boundary condition for the
continuum quasiparticle states using the HFB Green's function
$g_{LL'}$ described in the previous section. 

\begin{figure}[hptb]
  \begin{center}
    \includegraphics[width=8cm]{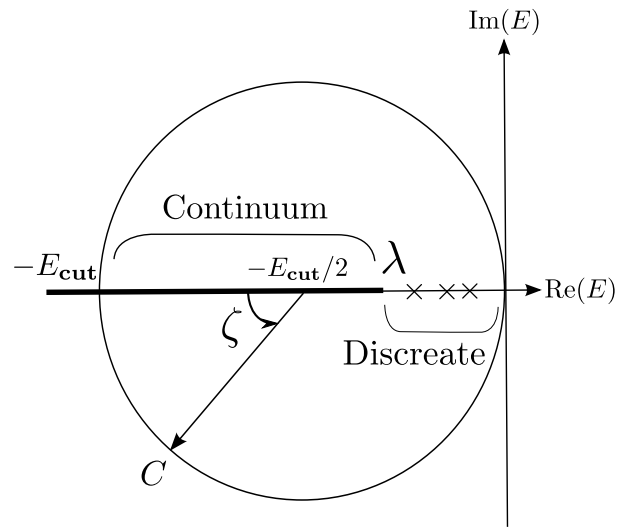}
     \caption{The circular path $C$ adopted to perform the
     contour integral (\ref{contour_int}). The thick solid line represents
     the negative energy continuum quasiparticle states
     with $E<\lambda$ while the crosses represent the
     negative energy discrete quasiparticle states $\lambda <E_i <0$.  
}\label{Complex_E_plane_c}
  \end{center}
\end{figure}

In order to make the numerical integral
efficiently, we choose $C$ as a circular path shown in Fig.\ref{Complex_E_plane_c} although
a rectangular contour is adopted in Ref.\cite{Green_sph_HFB}. Then
the integral reads
\begin{equation}
R_{LL'}(r,r') 
= \frac{E_{cut}}{4\pi i} \int_{0}^{2\pi}
g_{LL'}(r,r',E(\zeta))
ie^{i(\zeta-\pi)}d\zeta,
\label{contour_int}
\end{equation}
where 
$E_{cut}$ is the maximal quasiparticle energy which define the
cut-off of the sum $|E|<E_{cut}$ of the quasiparticle states. The circular path makes the integrand
to behave smoothly. 
In actual numerical calculations, we split $C$ into two semicircles 
$0< \zeta <\pi$
and $\pi < \zeta <2\pi$, and then  apply the higher-order Gauss-Legendre quadrature
with $M_\zeta/2$ points to perform the numerical integration in each semicircle 
($M_\zeta$ points in total). The local density $\rho(\vecr)=\sum_\sigma \rho(\vecr\sigma,\vecr\sigma)$
and the local pair density $\tilde{\rho}(\vecr)=\sum_\sigma \tilde{\rho}(\vecr\sigma,\vecr\sigma)$
are calculated accordingly.

Given the methods to calculate the quasiparticle wave functions and
the generalized density matrices, an usual iteration procedure can be
applied to obtain a converged HFB ground state $ | \Psi \ket$,
the associated quasiparticle states and the densities. The Fermi
energy $\lambda$ is also determined selfconsistently so that the
expectation value of the neutron/proton number is constrained to 
$N/Z$ of the nucleus under consideration.

\section{Application to a deformed drip-line nucleus}

In the following we shall illustrate the deformed continuum HFB
method with numerical examples.  We would like to discuss some 
features of the quasiparticle spectra and the pair correlation which
are characteristic to deformed nuclei near the neutron drip-line.
As an example, we choose $^{38}$Mg which is predicted to be
prolately deformed and close to the neutron drip-line 
in the mean-field 
calculations\cite{terasaki,lalazissis,stoitsov,stoitsov2,THO_table,Rodriguez,Gauss_HFB_2,FRDM}. 
Experimentally, $^{39}$Mg is unbound and $^{40}$Mg is the most neutron-rich bound isotope
identified so far\cite{MG40}. We do not deal with $^{40}$Mg 
for which the prediction of the 
deformation is more subtle
\cite{terasaki,lalazissis,stoitsov,stoitsov2,THO_table,Rodriguez,Gauss_HFB_2,FRDM}.

\subsection{Model and numerical procedure}

\begin{figure}[hptb]
  \begin{center}
     \includegraphics[width=8.5cm,origin=c]{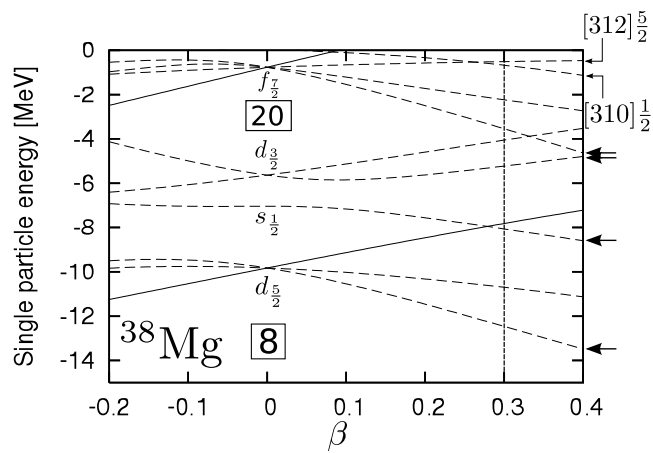}
     \caption{The neutron Nilsson diagram of the adopted
Woods-Saxon potential for $^{38}$Mg. The neutron Fermi energy is located near
the orbits labeled with 
the Nilsson asymptotic quantum numbers $[310]\frac{1}{2}$ and 
$[312]\frac{5}{2}$. The other orbits with $\Omega=1/2$ are also marked
with the arrows.
}\label{Mg38_Nilsson}
  \end{center}
\end{figure}

In the following analysis
we intend to clarify qualitative features rather than to make
 precise quantitative predictions of
the specific nucleus. We simplify the Hartree-Fock potential by replacing it with
an axially-symmetric deformed Woods-Saxon potential
\begin{equation}
V_{ws}(\boldsymbol{r}) = V_{ws}^{0}f(r,\theta) \label{def_ws},
\end{equation}
where $f(r,\theta)=(1+e^{-(r-R(\theta))/a_{0}})^{-1}$
and  $R(\theta)= R_{0}(1+\beta Y_{20}(\theta))$, together with
the spin-orbit potential
\begin{equation}
V_{ls}(\boldsymbol{r},\boldsymbol{s}) = 
V_{ls}^{0}\frac{r_{0}^2}{r}\left.\frac{df}{dr}\right|_{\beta=0} \vec{\ell} \cdot \vec{s}.
\end{equation}
We  choose the deformation parameter $\beta=0.3$, being a typical
value of the mean-field predictions
\cite{terasaki,lalazissis,stoitsov,stoitsov2,THO_table,Rodriguez,Gauss_HFB_2,FRDM}.  
The other Woods-Saxon parameters are
taken from Bohr and Mottelson\cite{Bohr_and_Mottelson}. 
We neglect the
deformation in the spin-orbit potential. The neutron Nilsson diagram of
the deformed Woods-Saxon potential is shown in Fig.\ref{Mg38_Nilsson}. If there is
no pair correlation, the highest occupied orbit is the one labeled with the
Nilsson asymptotic quantum number $[310]\frac{1}{2}$

Concerning the pairing properties, we derive it in a selfconsistent way
from an effective pairing interaction, for which we adopt the density-dependent
delta interaction (DDDI), 
\begin{equation}\label{dddi-int-eq}
v_\tau(\vecr-\vecr')=v_0 \frac{1-P_\sigma}{2}
 \left(1 - \eta \left({\rho_{\tau}(\vecr) \over
				     \rho_c}\right)^\alpha\right)\delta(\vecr-\vecr'). \ \ \
(\tau=n,p)
\end{equation}
The selfconsistent pair potential is then given by
\begin{equation}
\Delta(\boldsymbol{r}) =\frac{1}{2} v_{0} \left ( 1-\eta \left
					       (\frac{\rho_{\tau}(\boldsymbol{r})}{\rho_{c}}
					       \right )^{\alpha}
	\right ) \tilde{\rho}(\boldsymbol{r}). \label{pair_pot}
\end{equation}
Here we adopt the parameters ($v_0=-458.4\ {\rm MeV fm}^{-3}$, $\eta=0.76$, $\alpha=0.59$,
$\rho_c=0.08\ {\rm fm}^{-3}$) given in Ref.\cite{Matsuo07,Matsuo06}, where 
$v_0$ is determined so that the DDDI describes 
the $^1S$ scattering length $a=-18.5$ fm under the given cut-off energy $E_{cut}$.

The selfconsistent pair potential depends on 
the pair density 
$\tilde{\rho}(\boldsymbol{r})$, 
and it varies
in different physical situations. To provide a contrast to this case, 
we perform another calculation where the pair potential is replaced by
a fixed phenomenological one having a Woods-Saxon shape
\begin{equation}
\Delta(\boldsymbol{r}) = V_{p}^{0}f(r,\theta), 
\label{Analytic_volume}
\end{equation}
as adopted in Refs.\cite{hamamoto_def_HFB_1,hamamoto_def_HFB_2}.
According to Refs.\cite{hamamoto_def_HFB_1,hamamoto_def_HFB_2}
 we constrain the depth $V_{p}^{0}$ of the pair potential 
by the value of an average gap $\bar{\Delta}$ defined by 
$
\bar{\Delta} = {\int d \boldsymbol r \Delta(\boldsymbol r)
 f(r,\theta)}/
{\int d \boldsymbol r f(r,\theta)}. 
$

\begin{figure}[hptb]
  \begin{center}
     \includegraphics[width=7.5cm,origin=c]{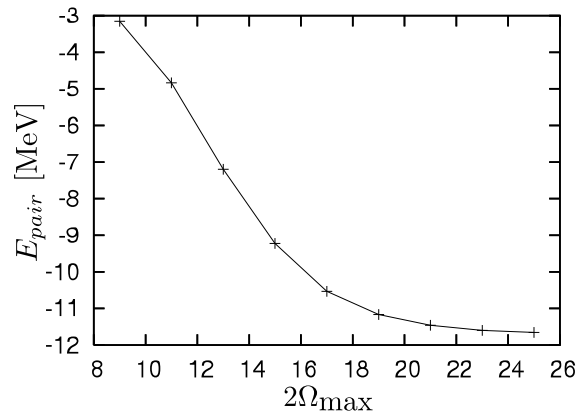} 
          \caption{Dependence of the neutron pair interaction energy $E_{pair}$ on
the maximal value $\Omega_{max}$ of the magnetic quantum number
of the quasiparticle states.
  }\label{dep_pair_energy}
  \end{center}
\end{figure}

The numerical details are as follows. We consider only neutrons.
Since we assume the axial symmetry,
the projection of the total angular momentum along the $z$-axis is conserved.
The quasiparticle states have the magnetic quantum number $\Omega$ , 
the eigenvalue of $j_z$,  and we solve the HFB equation separately for each $\Omega$.
The coupling potentials $u_{LL'}^{(0)}(r)$ and $\Delta_{LL'}^{(0)}(r)$
are evaluated according to Eqs.(\ref{dot_Vr}) and (\ref{dot_deltar}), where the integral over the
azimuthal angle $\theta$ is performed numerically by means of the
higher-order Gauss-Legendre quadruture with $N_\theta$ points. The
coupled-channel HFB equation (\ref{hfb_matrix}) is solved by means of the Runge-Kutta-Nyst\"{o}m
method\cite{RKN} in the region $r=0-R_{max}$ with $R_{max}=15$ fm and an equidistant
interval $\Delta r = 0.2$ fm. We impose the boundary condition that the
wave function is connected to the asymptotic form, Eq.(\ref{BC_asympt}), at
$r=R_{max}$. We use the code {\it cwfcomplex}\cite{pcwf} to calculate the Hankel and Coulomb
functions at complex energies.
The cut-off quasiparticle energy is $E_{cut}=60$ MeV. 
We also truncate the quasiparticle states in terms of the magnetic
number by including those with $-\Omega_{max} \le \Omega \le \Omega_{max}$.
Figure \ref{dep_pair_energy} shows the dependence of the neutron pair interaction energy
$E_{pair} = \frac{1}{2}\int d\vecr \Delta(\vecr)\tilde{\rho}(\vecr)$ 
on the choice of $\Omega_{max}$. A reasonable convergence is obtained
with $\Omega_{max} \gesim 21/2$, and hence  we adopt $\Omega_{max}=21/2$ in the following analysis.
The parameters for the Gauss-Legendre quadrature are $N_\theta=60$ and $M_{\zeta}=100$.

\subsection{Density profiles}

\begin{figure}[hptb]
  \begin{center}
    \begin{tabular}{c}
    \includegraphics[width=7.5cm]{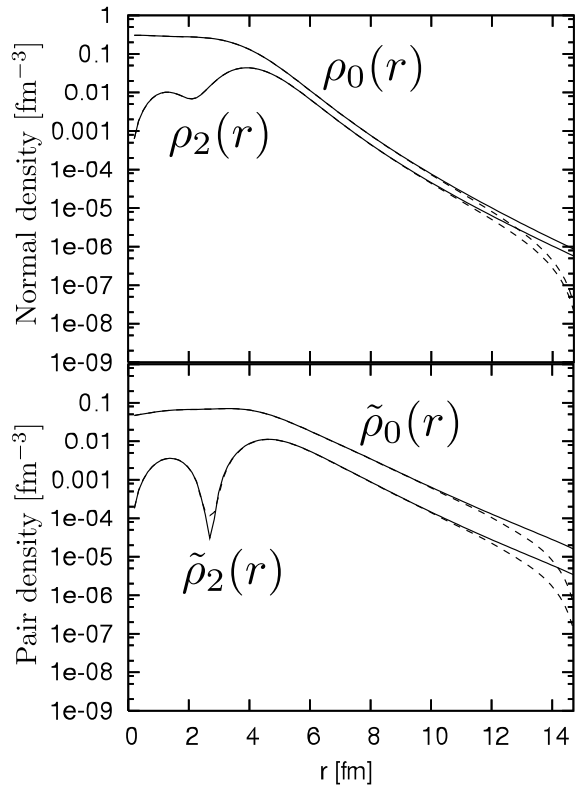}
    \end{tabular}
     \caption{(Top) The monopole and quadrupole parts 
$\rho_0(r)$ and $\rho_2(r)$ of the neutron density $\rho(\vecr)$.
The results of the calculation using the box boundary
condition are also plotted with the dashed curve for comparison.
(Bottom)  The same but for the monopole and quadrupole parts 
$\tilde{\rho}_0(r)$ and $\tilde{\rho}_2(r)$ of the neutron pair density 
$\tilde{\rho}(\vecr)$.  
}\label{n_dec}
  \end{center}
\end{figure}

The deformed continuum HFB method can describe the exponentially
decaying asymptotics of the densities. Figure \ref{n_dec} shows the
density profiles of the HFB ground state. Here we make the
multipole expansion of the densities as
\begin{eqnarray}
 \rho(\boldsymbol{r}) &=& \sum_{\ell} \rho_{\ell}(r) Y_{\ell 0}(\hat{r}), \\
 \tilde{\rho}(\boldsymbol{r}) &=& \sum_{\ell} \tilde{\rho_{\ell}}(r) Y_{\ell 0}(\hat{r}), 
\end{eqnarray}
and plot the monopole and quadrupole parts $\ell=0,2$.
For comparison, we also show the results of another calculation
where we impose the box boundary condition $\phi_L(R_{max})=0$ at the
boundary $r=R_{max}$. In the latter case the whole quasiparticle spectrum
including the continuum part
is discretized. It is clear that the calculation with the correct
boundary condition can describe the exponentially decaying asymptotics
both for the density and the pair densities while the calculation
with the box boundary condition fails.  

We remark here that both the monopole and quadrupole parts
exhibit the same slope in the exponentially decaying tail, i.e., 
the ratios $\rho_2(r)/\rho_0(r)$ and $\tilde{\rho}_2(r)/\tilde{\rho}_0(r)$ are
approximately independent of $r$ in the tail region.
This means that  we can define the ``deformation" of the exponentially decaying tails 
in terms of the shape of the equi-density contour curves. Note however that the 
size of the deformation in the tail differs from that of the nuclear surface (
which may be defined by the deformation of the
equi-density contour at the half central density). It is also seen that the
tail deformation in the density $\rho(\vecr)$ and that in
the pair density $\tilde{\rho}(\vecr)$ are different. It is interesting to
quantify the deformations in the tail region, but we leave it for future works
since in the present calculation we do not make the selfconsistent treatment of
the deformation of the Hartree-Fock potential.
The calculation in Ref.\cite{PTG_HFB} indicates also that the exponentially decaying tail
has the same slope in different directions.

\subsection{Weak-binding effect on pair correlation}\label{weakbinding_on_pairing}

We can regard the calculated $^{38}$Mg as
a quite weakly bound system since
the neutron Fermi energy in the present calculation is just $\lambda= -0.889$ MeV.
It is interesting to see what happens if the system is weakly bound further more,
or if the system is more strongly bound. To see the effects of the weak binding,
we investigate the pairing correlation by varying artificially the depth parameter
$V_{ws}^{0}$ 
of the Woods-Saxon potential as
\begin{equation}
 V_{ws}^{0} \rightarrow V_{ws}^{0}+\alpha
\end{equation}
where we choose $\alpha=+1,0,-1,-2,-3$ and $-4$ MeV.
The neutron Fermi energy is $\lambda= -0.471, -0.889, -1.335, -1.802, -2.289$ and $-2.793$ MeV
for $\alpha=+1,0,-1,-2,-3$ and $-4$ MeV, respectively.

\begin{figure}[hptb]
  \begin{center}
    \begin{tabular}{c}
     \includegraphics[width=8cm,origin=c]{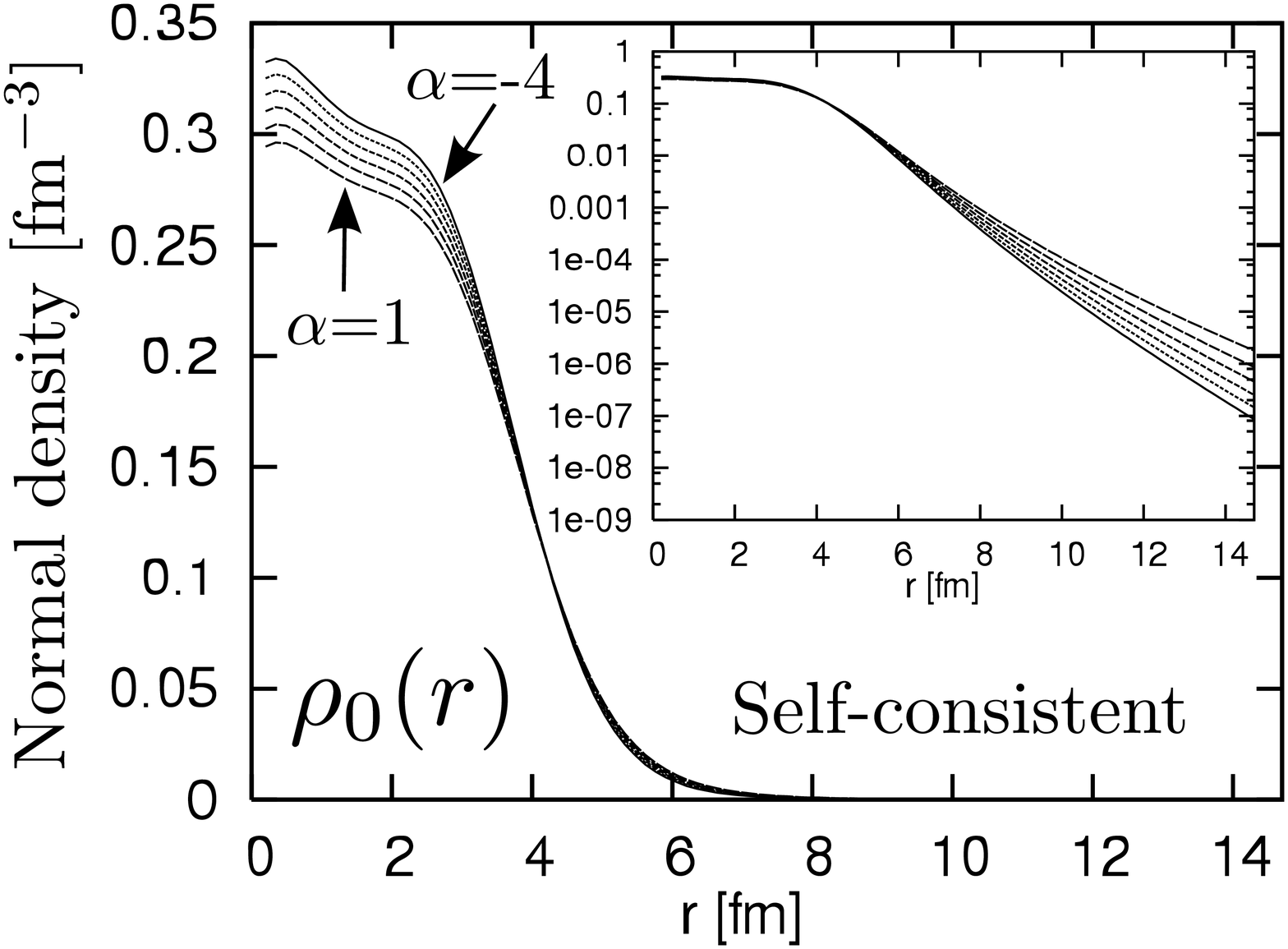} \\
\includegraphics[width=8cm,origin=c]{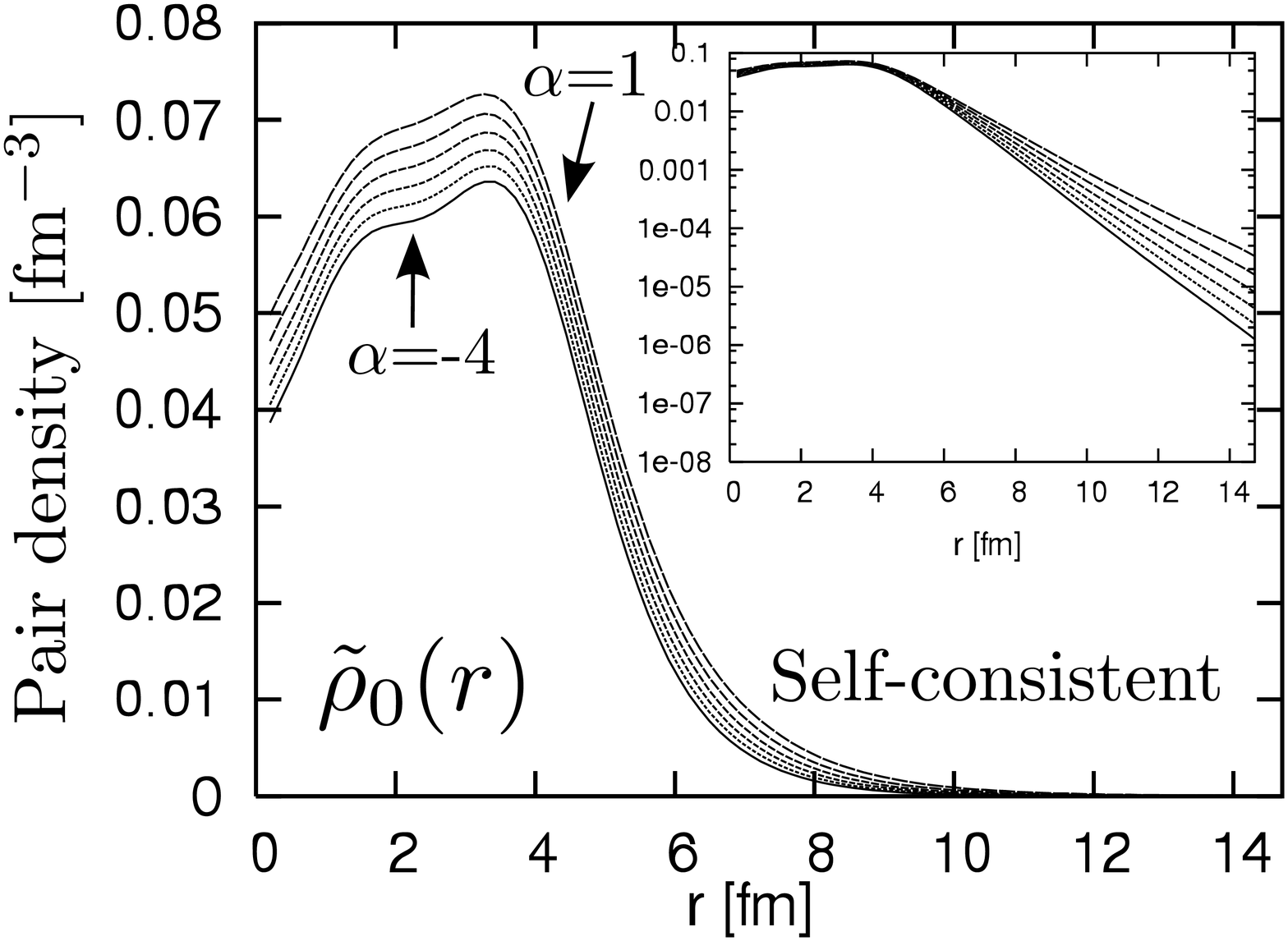}
    \end{tabular}
     \caption{(Top) The monopole part of the neutron density $\rho_0(r)$ for
     various shifts of the Woods-Saxon potential depth with $V_{ws}^{0} + \alpha$
     ($\alpha=+1,0,-1,-2,-3$ and $-4$ MeV). 
 The inset is the same plot but
in the log scale.
(Bottom) The same but for
 the monopole part of the neutron pair density $\tilde{\rho}_0(r)$.
}\label{n_dec_valV}
  \end{center}
\end{figure}

\begin{figure}[hptb]
  \begin{center}
    \begin{tabular}{c}
     \includegraphics[width=8cm,origin=c]{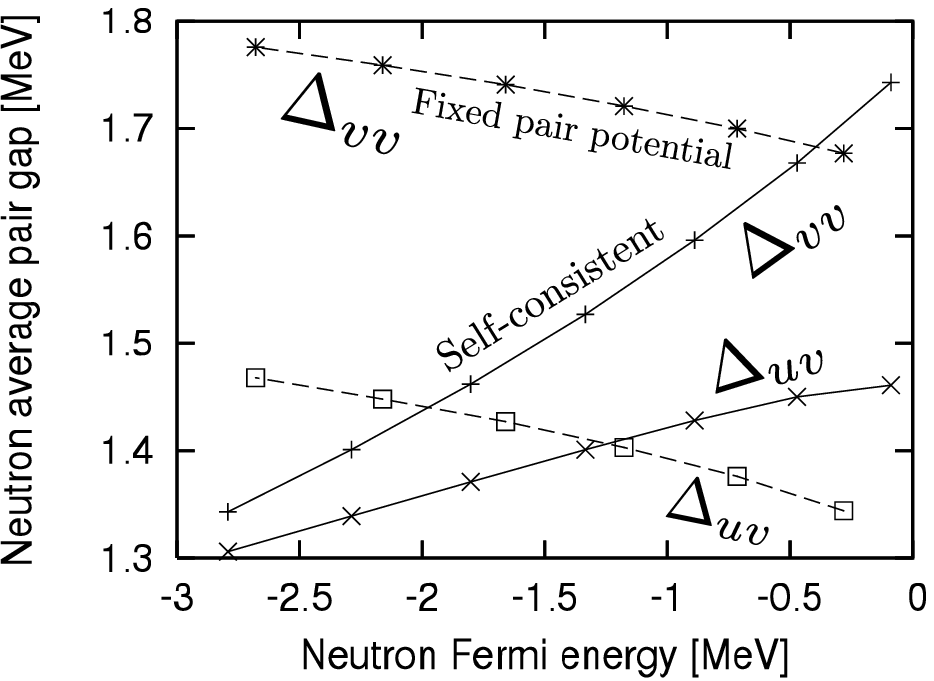} \\
     \includegraphics[width=8cm,origin=c]{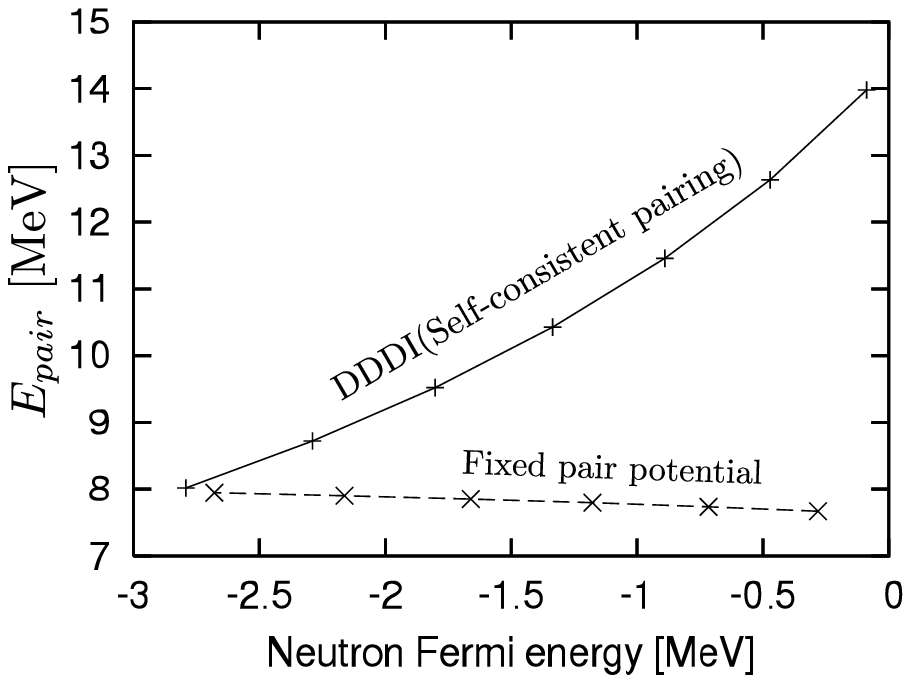}
    \end{tabular}
     \caption{(Top) The average pairing gaps $\Delta_{vv}$ and $\Delta_{uv}$ 
     plotted with the solid curves as a function of the neutron Fermi energy $\lambda$ for the
     selfconsistently calculated pair potential. The results for the fixed
     phenomenological pair potential  is also plotted with the dashed curves. 
    (Bottom) The     
     neutron pair interaction energy $E_{pair}$ 
    as a function of the neutron Fermi energy $\lambda$ 
    for the
     selfconsistently calculated pair potential (solid line) and also for the fixed
     phenomenological pair potential (dashed line). 
}\label{pair_gap_valV}
  \end{center}
\end{figure}

\begin{figure}[hptb]
  \begin{center}
    \begin{tabular}{c}
          \includegraphics[width=8cm,origin=c]{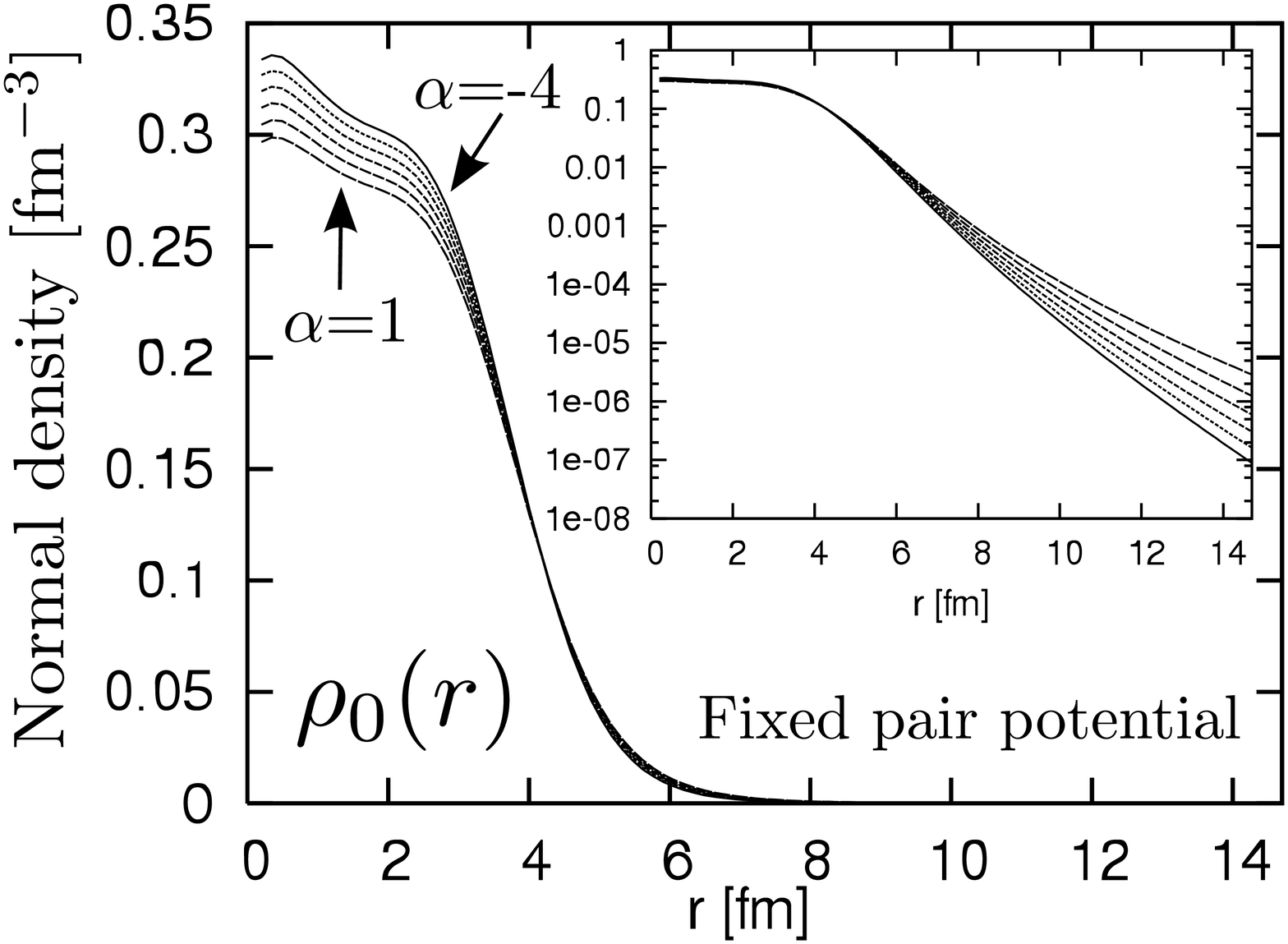} \\
\includegraphics[width=8cm,origin=c]{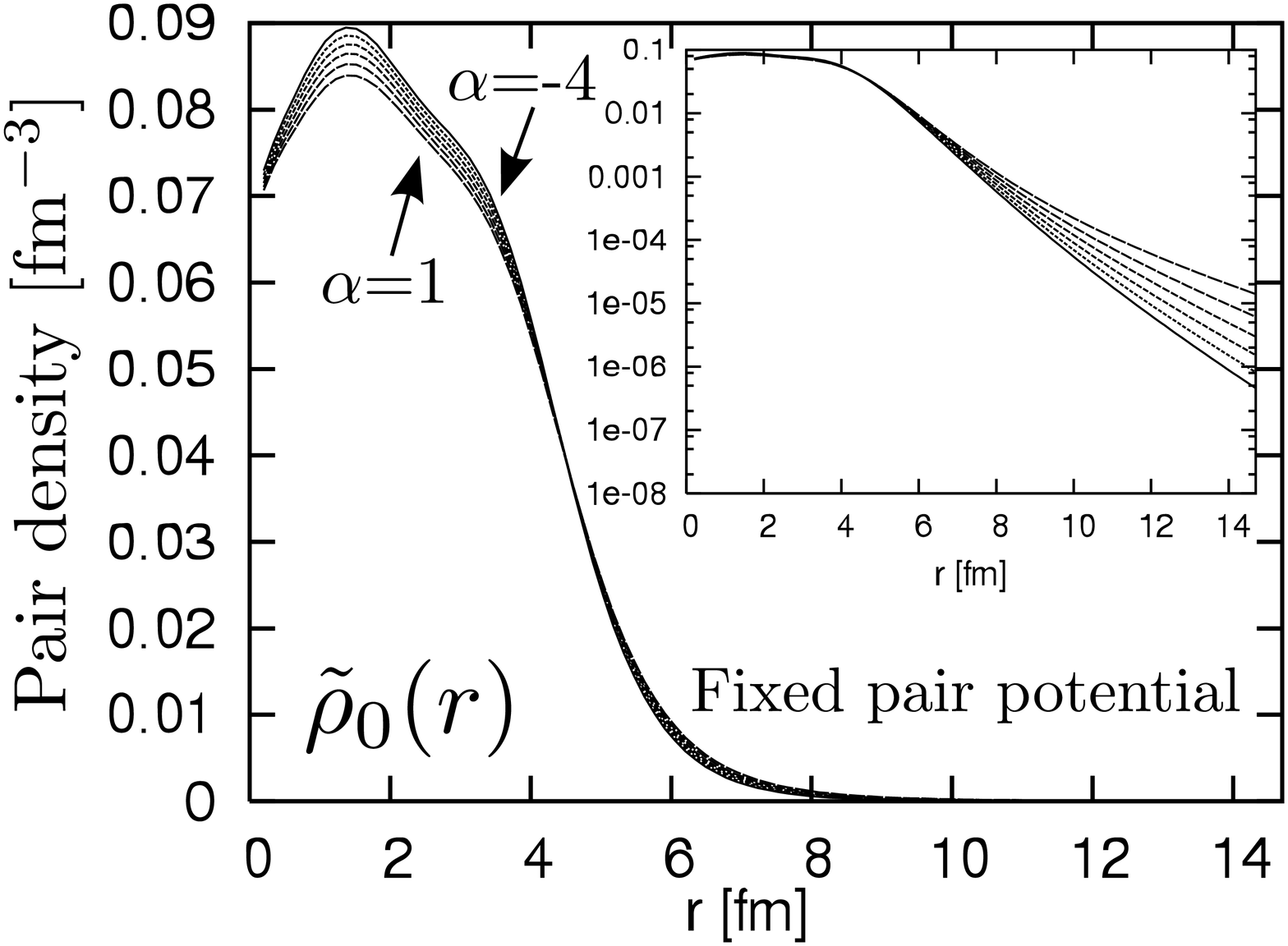}
    \end{tabular}
     \caption{ The same as Fig.\ref{n_dec_valV} but for the fixed phenomenological pair
     potential. 
   }\label{npd_l0_valV_con}
  \end{center}
\end{figure}

Figure \ref{n_dec_valV} shows the monopole parts $\rho_0(r)$ and $\tilde{\rho}_0(r)$ of
the density and pair density, respectively, of neutrons for different values of $\alpha$.
It is seen that the exponentially decaying tails of the density and the pair density
extends (shrinks) as the Fermi energy becomes shallower (deeper).  This is
naturally expected as the exponential decay behaves kinematically as
$\rho(r) \propto e^{-2\kappa r}$ and $\tilde{\rho}(r) \propto e^{-\kappa r}$ with
the exponent $\kappa=\sqrt{2m|\lambda|}/\hbar$ related to
 the Fermi energy $\lambda$\cite{DNW96}. 
We also see another noticeable
feature that as $\lambda$ becomes shallower the magnitude of the pair density increases 
not only in the tail region but also in the whole
region of $r$, including the interior and the surface areas. Apparently 
the pair correlation is enhanced by the weak binding. To
characterize the overall magnitude of the pair correlation we evaluate 
the 
average pairing gaps $\Delta_{vv}$ and $\Delta_{uv}$ defined by
\begin{equation}
\Delta_{vv} = \frac{\int d \boldsymbol r \Delta(\boldsymbol r)
 \rho(\boldsymbol r)}{\int d \boldsymbol r
 \rho(\boldsymbol r)},
\label{define_delta_vv}
\end{equation}
\begin{equation}
\Delta_{uv}= \frac{\int d \boldsymbol r \Delta(\boldsymbol r)
 \tilde{\rho}(\boldsymbol r)}{\int d \boldsymbol r
 \tilde{\rho}(\boldsymbol r)},
\label{define_delta_uv}
\end{equation}
and the pair interaction energy 
\begin{equation}
E_{pair}= \frac{1}{2}\int d\vecr \Delta(\vecr)\tilde{\rho}(\vecr).
\end{equation}
They are plotted in Fig.\ref{pair_gap_valV}, where
the enhancement due to the weak binding is seen directly in these
quantities.

In order to investigate the origin of the enhanced pairing in the
weak binding cases, we compare with another calculation where
the pair potential is replaced 
by a phenomenological one, Eq.(\ref{Analytic_volume}), with fixed strength
$\bar{\Delta}=1.5$ MeV.
The calculated
average pairing gaps $\Delta_{vv}$ and $\Delta_{uv}$
 and the pair interaction energy $E_{pair}$
are shown also in Fig.\ref{pair_gap_valV}, where, in contrast to the
selfconsistent case,  $\Delta_{vv}$ and $\Delta_{uv}$ stay
almost constant or even decreases slightly with lifting the
Fermi energy. 
The monopole density $\rho_0(r)$ and the monopole pair
density $\tilde{\rho}_0(r)$ are shown in Fig.\ref{npd_l0_valV_con}.
It is seen that the
 pair density  inside the nucleus 
decreases as the Fermi energy becomes shallower. This is opposite to the
trend seen in the case of the selfconsistent calculation.
Consequently 
there is very little influence of the weak binding on the pair 
correlation provided that we fix the pair potential.
In other words, the enhancement of the pair correlation caused
by the weak binding is described only when we calculate
the selfconsistent pair potential. In the selfconsistent case, the increase of
the pair density in the tail region induces the increase of the pair correlation
not only in the tail region but also in the entire region.

\begin{figure}[hptb]
  \begin{center}
    \begin{tabular}{c}
     \includegraphics[width=8cm,origin=c]{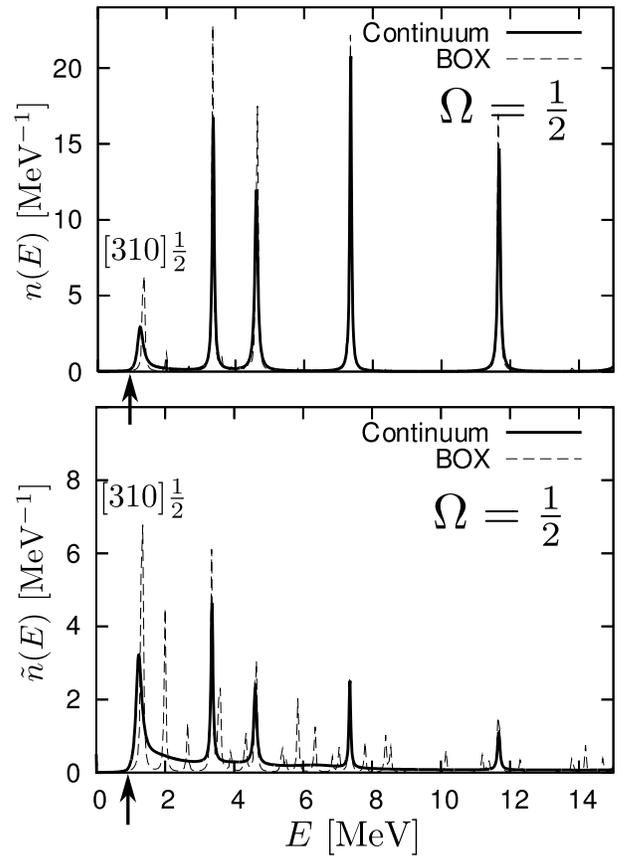}
    \end{tabular}
     \caption{The occupation number density $n(E)$ (top panel) and 
the pair number density $\tilde{n}(E)$ (bottom)
for the neutron quasiparticle states with $\Omega=1/2$.
The solid curve is the results of the deformed continuum HFB calculation with the
correct boundary condition on the asymptotics while
the dashed curve is obtained by imposing the box boundary condition $\phi(R_{max})=0$.
   The arrow indicates the threshold energy $|\lambda|=889$ keV 
for the continuum quasiparticle states. 
   }\label{level_CONvsBOX}
  \end{center}
\end{figure}

\subsection{Structure of continuum quasiparticle states}

In the HFB theory, both the neutron density $\rho(\vecr)$  and 
the expectation value $\left<N \right> =\int d\vecr \rho(\vecr)$ of the neutron number can be
expressed as summations of contributions of the individual quasiparticle states. Since the
quasiparticle states are specified by the energy eigenvalue $E$, we can define 
the contributions per unit energy 
(the density per energy) to the neutron number $\left< N \right> $. We denote it by $n(E)$,
and call the occupation number density. It
satisfies $\left< N \right>=\int_0^{E_{cut}} dE n(E)$ by definition, and can be
calculated in terms of the HFB Green's function by
\begin{equation}
n(E)=\frac{1}{\pi} \mbox{Im} \sum_{\sigma}\int
 d\boldsymbol{r}G^{(11)}(\boldsymbol{r}\sigma,\boldsymbol{r}\sigma,-E-i\epsilon).
\label{nn_den}
\end{equation}
This is the level density of the quasiparticle states, but it is
weighted with the occupation 
number (corresponding to the $v^2$ factor in the case of the BCS approximation).
Similarly we can define the pair number density $\tilde{n}(E)$
by 
\begin{equation}
\tilde{n}(E)=\frac{1}{\pi} \mbox{Im} \sum_{\sigma}\int
 d\boldsymbol{r}G^{(12)}(\boldsymbol{r}\sigma,\boldsymbol{r}\sigma,-E-i\epsilon)
\label{pn_den}
\end{equation}
for which the level density is weighted with the pairing factor (the $uv$-factor in the BCS). 
$\tilde{n}(E)$ represents the contribution of the quasiparticle state at energy $E$ 
to the pair number $\left< \tilde{N} \right> =\int d\vecr \tilde{\rho}(\vecr)$.
Note that the constant $\epsilon$ plays a role of the smoothing parameter.
If the quasiparticle spectrum contains discrete states, there emerge
 delta functions in $n(E)$ and $\tilde{n}(E)$ 
at the energies of 
the discrete quasiparticle states. Having the  imaginary part $i\epsilon$ 
in Eqs.(\ref{nn_den}) and (\ref{pn_den}),
the delta function  becomes a Lorentzian function with FWHM of $\gamma=2\epsilon$.
In the following, we take $\epsilon=25$ keV ($\gamma=50$ keV), and 
we shall investigate the structure of the quasiparticle spectrum
in terms of $n(E)$ and $\tilde{n}(E)$.

Figure \ref{level_CONvsBOX} shows  $n(E)$  and $\tilde{n}(E)$
for neutron states with $\Omega=1/2$. Note that
the threshold energy of the continuum spectrum is $E_{th}=|\lambda|=889$ keV.
It is seen that the continuum spectrum above $E_{th}$
 consists of several resonances forming narrow peaks, and a broad 
continuum background for which we hardly assign resonance structures.
We can identify a resonance at $E=1.24$ MeV as
a quasiparticle state corresponding to the orbit $[310]\frac{1}{2}$ of the
deformed Woods-Saxon potential (cf. Fig.\ref{Mg38_Nilsson}). 
Narrow resonances located at $E=3.36, 4.61, 7.36$ and 11.68 MeV correspond to
the other $\Omega=1/2$ orbits marked with the arrows  
in Fig.\ref{Mg38_Nilsson}.
The widths of these resonances are 
smaller than the smoothing width $\gamma=50$keV except the lowest energy 
resonance at $E=1.24$ MeV, 
which has a width apparently larger than those of the others.
Note that the $[310]\frac{1}{2}$ orbit is a bound state
if we neglect the pair potential, and 
the finite and relatively large width is brought by the pair potential.

Another important feature seen in Fig.\ref{level_CONvsBOX} is 
that the non-resonant continuum states  have significant
contribution to the pair number density $\tilde{n}(E)$
in an extent comparable to that
of the narrow resonance states. 
This should be contrasted to that the contributions of 
the non-resonant continuum states to 
the occupation number density $n(E)$ is very small. The small contribution of
the non-resonant continuum states to the occupation number density $n(E)$
is pointed out in Ref.\cite{hamamoto_def_HFB_2}, but we emphasize here that 
this is not the case for
the pair number density $\tilde{n}(E)$.

In Fig. \ref{level_CONvsBOX},
we also show the results obtained with the box boundary condition $\phi(R_{max})=0$.
In this case the quasiparticle states are all discretized. It is obvious that
the discretized quasiparticle states obtained with the box size $R_{max}=15$ fm
fail to describe the width of the resonances, for instance, that of the
$[310]\frac{1}{2}$. Obviously the discretization is too crude to 
describe the continuous behavior of the non-resonant
 continuum states.

\begin{figure}[hptb]
  \begin{center}
    \begin{tabular}{c}
    \includegraphics[width=8cm,origin=c]{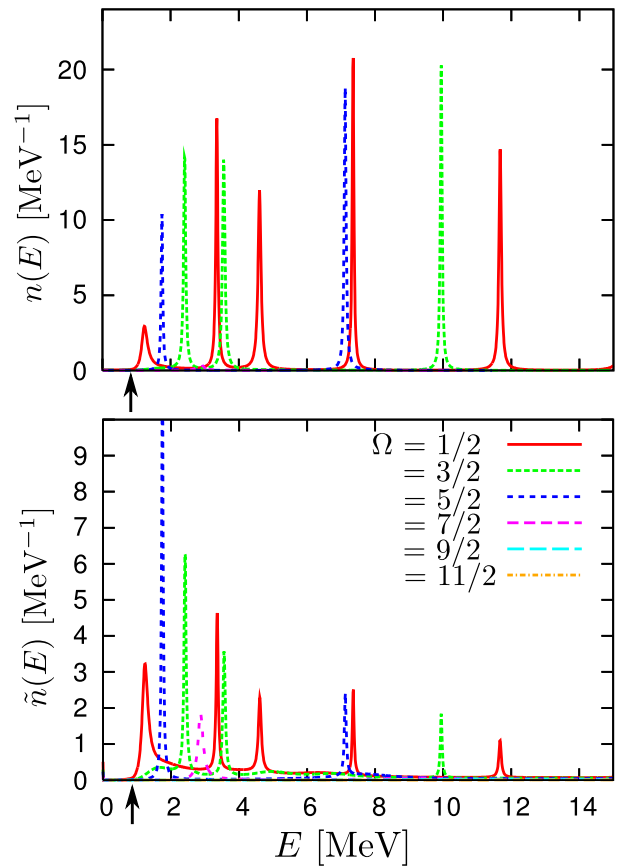} 
    \end{tabular}
     \caption{(Top) The occupation number density $n(E)$ of neutrons shown
   separately for different values of $\Omega=1/2,3/2,\cdots,11/2$. 
   The arrow indicates the threshold energy $|\lambda|$ for the continuum quasiparticle states.
   (Bottom) The same but for the pair number density $\tilde{n}(E)$.
}\label{spectrum_all}
\end{center}
\end{figure}

\begin{figure}[hptb]
  \begin{center}
    \begin{tabular}{c}
         \includegraphics[width=8cm,origin=c]{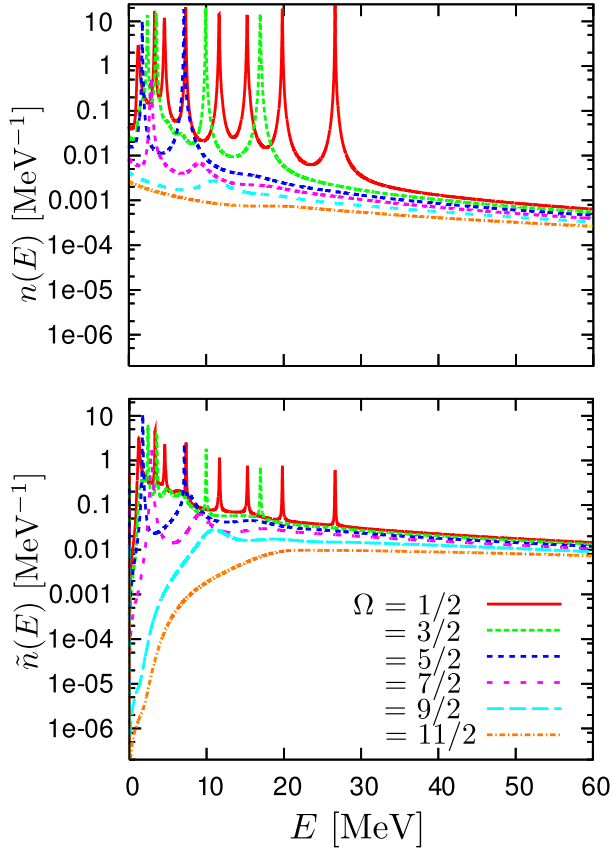} 
    \end{tabular}
     \caption{The same as Fig.\ref{spectrum_all} but in the log scale and in a wider energy range.
}\label{spectrum_all_log}
  \end{center}
\end{figure}

In Figs. \ref{spectrum_all} and \ref{spectrum_all_log} 
we show  $n(E)$ and $\tilde{n}(E)$
combining several values of $\Omega$.
We mention here two features. (i) Firstly, 
the lowest energy resonance (with $\Omega=1/2$) and 
the second lowest one (with $\Omega=5/2$) have significantly
different widths even though the centroid energies of both resonances are
comparable.
We note here that the single-particle energies of the orbits $[310]\frac{1}{2}$ 
and $[312]\frac{5}{2}$, which correspond to the lowest and the second lowest 
resonances, respectively, are also very close (cf. Fig.\ref{Mg38_Nilsson}).
The difference in the width may be attributed to the difference in the 
magnetic quantum number $\Omega$ in a way discussed in 
Ref.~\cite{hamamoto_def_WS}: the centrifugal barrier
for the $[310]\frac{1}{2}$ state is lower than that of the 
$[312]\frac{5}{2}$ state since the former 
contains the partial wave with $l=1$ while
the latter does not.
(ii) Secondly, it is seen
in the pair number density $\tilde{n}(E)$ that
the contribution of
the non-resonant continuum states is larger for states with smaller $\Omega$. 
 Note, however, that
the relative importance of high/low-$\Omega$ non-resonant states depends on
the quasiparticle energy. 
The low-$\Omega$ states such as
$\Omega=1/2$ and $3/2$ are dominant as far as the non-resonant quasiparticle states at
low energies $E\lesim 10$ MeV are concerned (cf. Fig.\ref{spectrum_all}). 
Concerning the 
non-resonant continuum states at higher energies $E\gesim 30$ MeV, however,
we see in Fig.\ref{spectrum_all_log} that 
contributions from various $\Omega$'s do not depend very strongly on
$\Omega$:   
there is only a small difference by a factor of $\sim 2$ in $\tilde{n}(E)$ 
between $\Omega=1/2$ and $11/2$ states at $E\gesim 30$ MeV. In other words, the 
high-$\Omega$ states also have sizable contributions to the pair correlation.

\begin{figure}[hptb]
  \begin{center}
    \begin{tabular}{c}
    \includegraphics[width=9cm,origin=c]{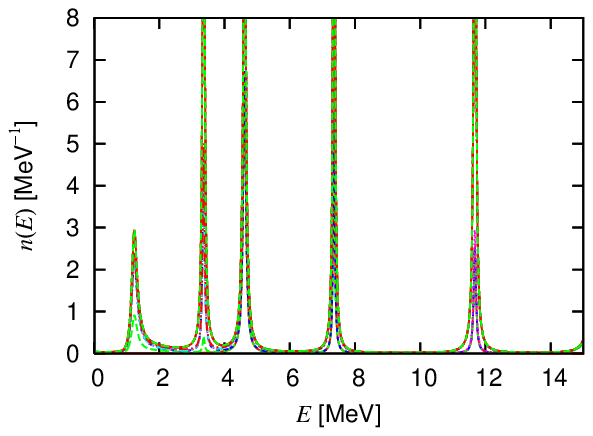} \\
  \includegraphics[width=9cm,origin=c]{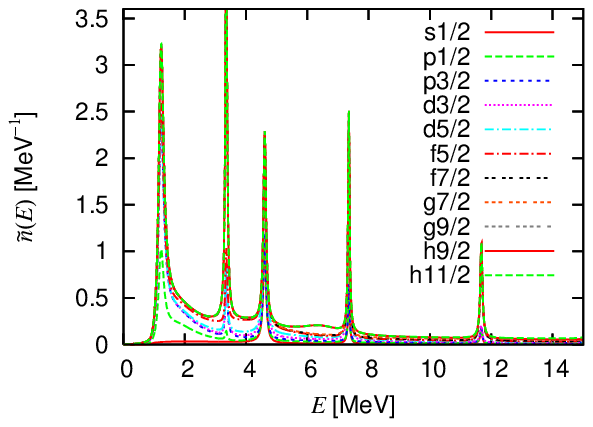}
    \end{tabular}
     \caption{
   The partial wave decomposition of the occupation number density $n(E)$ (top panel) and
   the pair number density $\tilde{n}(E)$ (bottom) of neutron $\Omega=1/2$ states.
   Each curve labeled with $lj$ represents 
the partial sum $\sum_{L<lj}n_L(E)\ \left(\tilde{n}_L(E)\right)$ 
containing from $s_{1/2}, p_{1/2}, p_{3/2}, 
   \cdots$ to $lj$. 
   }\label{level_par_dec}
  \end{center}
\end{figure}

The decomposition of the quasiparticle states into the partial waves is another 
useful tool to analyze the structure of the quasiparticle states.
In fact we can decompose the occupation number density $n(E)$ and 
the pair number density $\tilde{n}(E)$ into
different partial waves: 
\begin{eqnarray}
n(E) = \sum_{L} n_{L}(E), \ \ \ \ 
\tilde{n}(E) = \sum_{L} \tilde{n}_{L}(E),
\end{eqnarray}
where the contribution from each partial wave $L$ is given by
\begin{eqnarray}
n_{L}(E) = \frac{1}{\pi} \mbox{Im} \int  dr 
r^{2}g^{(11)}_{LL}(r,r,-E-i\epsilon)
\end{eqnarray}
and similarly for $\tilde{n}_{L}(E)$.
The partial wave decomposition of the $\Omega=1/2$ states is shown
in Fig.\ref{level_par_dec}.
It is seen that each of the resonances has specific partial wave
contents reflecting the character of the resonances. For instance, the 
lowest energy resonance $[310]\frac{1}{2}$
consists of  dominant $p_{1/2}-p_{3/2}$ components and
sub-dominant $f_{7/2}$ component. The second resonance consists of
dominant $f_{7/2}$ and sub-dominant $p_{3/2}$. 

We can also argue the structure of the non-resonant quasiparticle states
in terms of the partial wave
contents in $\tilde{n}(E)$. For instance, the continuum states in the
interval $1.5 \lesim E \lesim 3.0$ MeV is characterized by 
the relatively largest component of $p_{1/2}-p_{3/2}$  waves 
and sub-dominant $f_{7/2}-f_{5/2}$ components. 
We see also some structures in 
the intervals $2.5 \lesim E \lesim 5.5$ MeV and $5.5 \lesim E \lesim 8$,
which are characterized the relatively largest components
$f_{5/2}$   and  $g_{9/2}$, respectively. These structures 
presumably originate from resonance states in the deformed Woods-Saxon potential
lying high above the Fermi energy. They
coexist in the same energy region with smooth (almost flat in $E$) components containing
several partial waves with $l=0-4$. At higher energies, say $E\gesim 30$ MeV,
there remain only smooth non-resonant components 
including all the partial waves up to very high orbital angular momentum 
$l \lesim 6-10$. 
The contributions of high-$l$
orbits is consistent with the high-$\Omega$ components observed in 
Fig.\ref{spectrum_all_log}.

\subsection{Weak-binding effects on quasiparticle motion}

In this subsection we shall elucidate how the weak binding influences the
neutron quasiparticle motion and its coupling to the
pair correlation. For this purpose we first analyze
how   
the neutron pair number
density $\tilde{n}(E)$ evolves  when we vary the
Fermi energy by shifting the depth $V_{ws}^{0}$ of the Woods-Saxon potential as
$V_{ws}^{0} \rightarrow V_{ws}^{0} + \alpha$.

\begin{figure}[hptb]
  \begin{center}
\includegraphics[width=8.5cm]{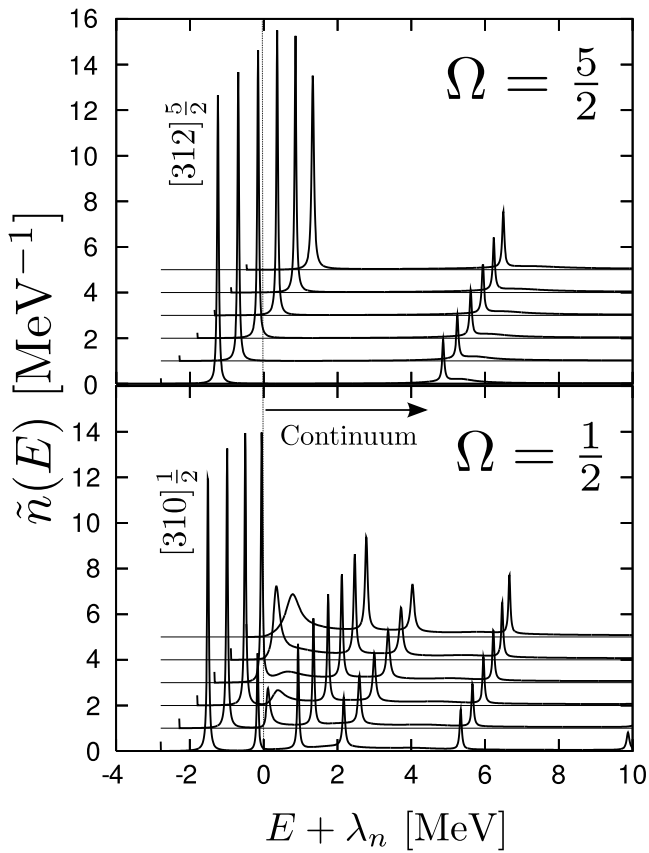} 
  \end{center}
  \caption{(Bottom) The pair number density $\tilde{n}(E)$ of the neutron $\Omega=1/2$
  states is shown for various shifts $\alpha=-4, -3, -2, -1, 0$ and +1 MeV 
  of the Woods-Saxon potential depth
  $V_{ws}^{0} \rightarrow V_{ws}^{0}+\alpha$. The horizontal axis represents the
  quasiparticle energy $\lambda+E$ whose origin corresponds to
 the threshold energy $E_{th}=|\lambda|$
  for the continuum spectrum.
 (Top) The same but for the
  $\Omega=5/2$ states.
}
\label{spectrum_varWS}
\end{figure}

\begin{figure}[hptb]
  \begin{center}
   \includegraphics[width=8cm,origin=c]{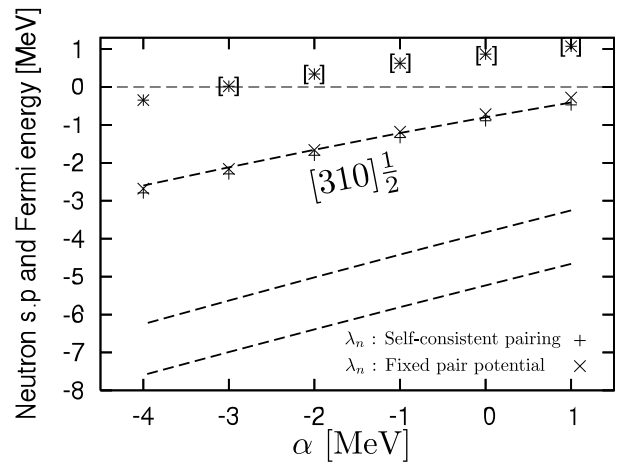} 
     \caption{The single-particle energy of the neutron $\Omega=1/2$ orbits in
the deformed Woods-Saxon potential without the pair correlation
for various shifts $\alpha=-4, -3, -2, -1, 0$ and +1 MeV 
  of the Woods-Saxon potential depth
  $V_{ws}^{0} \rightarrow V_{ws}^{0}+\alpha$. The positive
energy states are discretized with the box boundary condition at $r=R_{max}=15$fm,
and shown within the parenthesis. The calculated neutron Fermi energy $\lambda$ is 
plotted also with the symbol $+$. $\lambda$ calculated for the fixed phenomenological pair potential
is also plotted with the symbol $\times$.
   }\label{WS_sp}
  \end{center}
\end{figure}

Figure \ref{spectrum_varWS} shows 
the pair number density $\tilde{n}(E)$ of 
$\Omega=1/2$ and $5/2$ states calculated 
for various Woods-Saxon potential depths. 
It is seen first of all that the influence of the weak binding is stronger on the
 $\Omega=1/2$ states than on the $\Omega=5/2$ states. 
(This is along with the observation in Refs.
\cite{hamamoto_def_HFB_1,hamamoto_def_HFB_2,hamamoto_def_WS}.)
The influence is especially strong on the low-lying quasiparticle
 states. Here we concentrate on $\Omega=1/2$, and focus on 
the lowest energy
quasiparticle state (or the quasiparticle resonance) which corresponds to the
$[310]\frac{1}{2}$ orbit, the second lowest one that is visible only
for $\alpha=-4,-3$ and $-2$ MeV (and may be also $\alpha=-1$) cases,  and the 
non-resonant structures around these resonance states. 
Note that there are little influences of the weak binding on the
 third, fourth and fifth lowest peaks, which correspond to the hole orbits deeper than the
 Fermi energy by $3 \sim 8$ MeV (cf. Fig.\ref{Mg38_Nilsson}).

Let us look into the $[310]\frac{1}{2}$ state. 
The binding energy of the $[310]\frac{1}{2}$ orbit of the 
deformed Woods-Saxon potential becomes very small 
in the cases of $\alpha=0$ and $+1$ MeV
(cf. Fig.\ref{WS_sp}).  In these cases the quasiparticle peak 
corresponding to  the $[310]\frac{1}{2}$ orbit  is 
located in the continuum region $E+\lambda > 0$ ($E>|\lambda|= E_{th}$).
It has a finite width due to the
the coupling to the continuum.
 It is seen that the width increases quite steeply for
the variation of the potential depth $\alpha=-1 \rightarrow 0 \rightarrow +1$ MeV.
The steep change in the width is quite contrasting to that of the $\Omega=5/2$ case,
i.e. the $[312]\frac{5}{2}$ state,
for which there is only a small change in the width.
Note also that
 a similar steep increase of the width can be seen in the second lowest peak, 
which corresponds to the Woods-Saxon orbit marked with the asterisk in Fig. \ref{WS_sp}.
The second lowest peak is a bound discrete state 
in the case of $\alpha=-4$ MeV, and once it becomes unbound resonance for 
$\alpha\ge -3$ MeV the width increases steeply along with the variation
 $\alpha=-3 \rightarrow -2 \rightarrow -1$ MeV.
With $\alpha=-1$ MeV we 
see only a broad distribution or a wide resonance,
whose width could be order of 1 MeV. With $\alpha= 0$ and $+1$ MeV, 
there remains only a smooth continuum which is hard to be identified
as a resonance structure.

The most  important feature which we observe in Fig.\ref{spectrum_varWS}
 is that
the contribution of the low-lying non-resonant continuum quasiparticle
states to $\tilde{n}(E)$ increases significantly as the neutron Fermi energy
becomes close to zero, especially in 
the cases $\alpha=0$ and $+1$ MeV (corresponding to $\lambda=-0.89$ and $-0.47$ MeV,
respectively).
If we look closely at the second lowest resonance discussed above,  it
keeps  a sizable contribution to $\tilde{n}(E)$ in the cases $\alpha=-2$ and $-1$ MeV
where the width becomes very large. The contribution is still
sizable even in the cases $\alpha=0$ and $+1$ MeV where only a 
 non-resonant continuum structure is visible as a remnant in the expected energy region. 
It is argued in Refs.\cite{hamamoto_def_HFB_1,hamamoto_def_HFB_2} that 
weakly bound or resonant $\Omega=1/2$ 
quasiparticle states contribute little to the pair correlation compared with that of
the strongly bound states. Our calculation does not exhibit this behavior.
Our analysis rather suggests that all of the weakly bound, resonant
and non-resonant quasiparticle states contribute significantly to the pair correlation
in the case when the neutron Fermi energy is small. This is also supported by
the observation in subsection~\ref{weakbinding_on_pairing} 
that the weak binding enhances the net pair correlation.

\begin{figure}[hptb]
  \begin{center}
    \begin{tabular}{c}
         \includegraphics[width=8cm,origin=c]{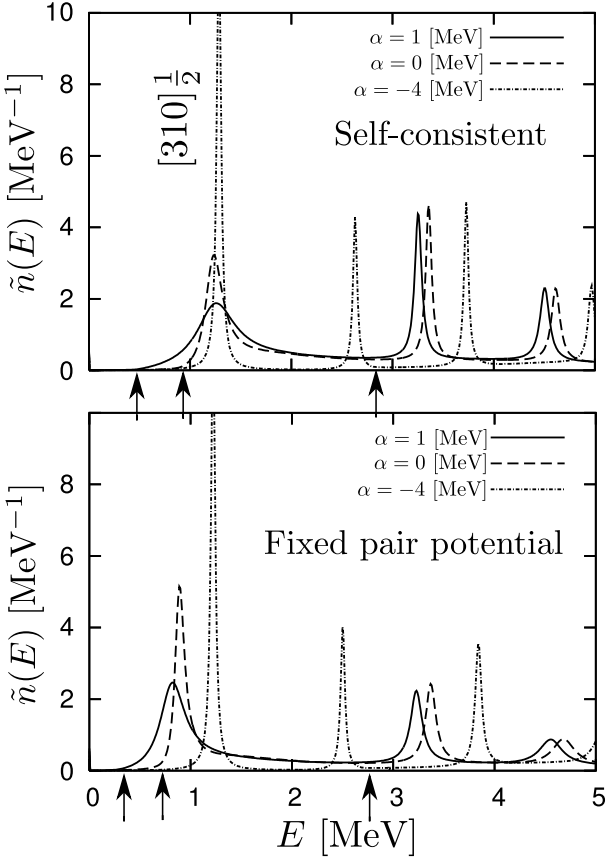}
    \end{tabular}
  \end{center}
     \caption{(Top) The pair number density $\tilde{n}(E)$ of the neutron $\Omega=1/2$
  states for the shifted Woods-Saxon potential depths 
$V_{ws}^{0} \rightarrow V_{ws}^{0}+\alpha$ with $\alpha= +1, 0$ and $-4$ MeV.
  The arrow indicates
  the positions of the Fermi energy.
  This is the result of  the calculation using the selfconsistent pair potential derived from
  the DDDI. (Bottom) The same but for the calculation assuming the fixed phenomenological pair potential.
"Selfconsistent"'Æ"Fixed pair potential"'͘gŠO'Å'Í'È'­˜g"à'É'¢'ê'éB$\alpha=0,-4$'ɈႤüŽí'ðŽg'¤B}
\label{spectrum_varWS_SCvsanalytic}
\end{figure}

\begin{table}[h]
\caption{The peak energy $E_{peak}$ of the discrete quasiparticle states and resonances
in the pair number density $\tilde{n}(E)$ for $\Omega=1/2$, and the effective pairing gaps 
$\Delta_{uv}(E_{peak})$ 
and $\Delta_{vv}(E_{peak})$ of the corresponding quasiparticle states, 
calculated for the 
shifted Woods-Saxon potential depths $V_{ws}^{0}+\alpha$ with $\alpha=+1,0, -4$ MeV.
The unit is MeV for all the quantities. In the upper half is listed the results obtained with
the selfconsistent pair potential derived from the DDDI while the lower half is for
those obtained with the fixed phenomenological pair potential, Eq.(\ref{Analytic_volume}).}
 \label{gap_value_SC}
\begin{center}
\begin{tabular}{ccc @{\hspace{15pt}} ccc @{\hspace{15pt}} ccc}\hline\hline
\multicolumn{3}{c}{$\alpha=1$  } &
\multicolumn{3}{c}{$\alpha=0$ } &
\multicolumn{3}{c}{$\alpha=-4$  } \\
$E_{peak}$ & $\Delta_{vv}$ & $\Delta_{uv}$  &
$E_{peak}$ & $\Delta_{vv}$ & $\Delta_{uv}$ &
$E_{peak}$ & $\Delta_{vv}$ & $\Delta_{uv}$ \\ \hline
\multicolumn{9}{l}{Selfconsistent pair potential} \\
1.26 & 1.39 & 1.14 & 1.24 & 1.39 & 1.19 & 1.28   & 1.30 & 1.27  \\ 
3.25 & 1.71 & 1.64 & 3.36 & 1.64 & 1.59 & 3.73  & 1.41 & 1.37  \\ 
4.50  & 1.62 & 1.62 & 4.61  & 1.55 & 1.54 & 4.97  & 1.30 & 1.25  \\ 
7.14 & 1.74 & 1.65 & 7.36 & 1.66 & 1.56 & 8.14  & 1.42 & 1.26  \\ 
11.39 & 1.71 & 1.48 & 11.68 & 1.62 & 1.42 & 12.67 & 1.35 & 1.14 \\ \hline

\multicolumn{9}{l}{Fixed pair potential} \\
0.82   & 1.00 & 0.72 & 0.89   & 1.08 & 0.84 & 1.22  & 1.32 & 1.21  \\ 	  
3.23  & 1.45  & 1.08  & 3.37  & 1.48 & 1.11  & 3.84 & 1.58 & 1.25  \\ 	
4.56  & 1.41  & 1.11  & 4.68  & 1.44 & 1.07  & 5.07 & 1.53 & 1.03 \\ 	
7.32   & 1.67  & 1.72  & 7.55  & 1.69 & 1.75  & 8.34 & 1.78 & 1.83  \\ 	
11.66 & 1.81  & 2.41  & 11.94 & 1.83 & 2.42  & 12.90 & 1.88 & 2.42  \\ \hline\hline
\end{tabular}
\end{center}
\end{table}

Let us then investigate 
how the weakly bound and resonant quasiparticle states couple with 
the pair correlation. 
For this purpose we shall analyze the pairing gap associated with these
orbits. As far as the $[310]\frac{1}{2}$ state is concerned,
we can evaluate an effective pairing gap of this state in a simple way.
This is because, as seen
 in Fig.\ref{WS_sp}, the neutron Fermi energy $\lambda$ coincides within 0.1 MeV with 
the deformed Woods-Saxon single-particle energy $e_{i}$ of the $[310]\frac{1}{2}$ orbit 
for all the values of $\alpha=-4,\cdots,0,1$ MeV. In this situation, the quasiparticle
energy $E_{i}$ of this state can be regarded\cite{hamamoto_def_HFB_1} 
as an effective pairing gap $\Delta_i^{(eff)}$ 
since the approximate relation
\begin{equation}
 E_{i}=\sqrt{(e_{i}-\lambda)^{2}+(\Delta_{i}^{(eff)})^{2}} \label{BCS_gap}
\end{equation}
valid in the BCS approximation leads to 
$E_i \approx \Delta_i^{(eff)}$ for
$|e_i -\lambda| \ll \Delta_i^{(eff)}$.
We read from Fig. \ref{spectrum_varWS_SCvsanalytic} 
and Table \ref{gap_value_SC} that the peak position of the $[310]\frac{1}{2}$
resonance stays at $E_{i} \approx 1.3$ MeV, and hence $\Delta_{i}^{(eff)}\approx 1.3$ MeV
in all the cases. Namely we see that
the effective pairing gap is almost unchanged even when $e_i \rightarrow 0$.

We can evaluate more directly effective paring gap associated with the quasiparticle states.
Note here that  the density $\rho(\vecr,E)$  and the pair density $\tilde{\rho}(\vecr,E)$
associated with the quasiparticle state at $E$ is given by 
\begin{eqnarray}
\rho(\vecr,E) &=&  \frac{1}{\pi} \mbox{Im} \sum_{\sigma}\int 
d\boldsymbol{r}G^{(11)}(\boldsymbol{r}\sigma,\boldsymbol{r}\sigma,-E-i\epsilon), \\
\tilde{\rho}(\vecr,E) &=& \frac{1}{\pi} \mbox{Im} \sum_{\sigma}\int
 d\boldsymbol{r}G^{(12)}(\boldsymbol{r}\sigma,\boldsymbol{r}\sigma,-E-i\epsilon).
\end{eqnarray}
Utilizing these quantities, we
can evaluate the effective paring gaps of this state by 
\begin{eqnarray}
\Delta_{vv}(E) &=& \frac{\int d \boldsymbol r \Delta(\boldsymbol r)
 \rho(\boldsymbol r,E)}{\int d \boldsymbol r
 \rho(\boldsymbol r,E)}, \\
\Delta_{uv}(E) &=& \frac{\int d \boldsymbol r \Delta(\boldsymbol r)
 \tilde{\rho}(\boldsymbol r,E)}{\int d \boldsymbol r
 \tilde{\rho}(\boldsymbol r,E)}.
\end{eqnarray}
 The calculated values of the
effective pairing gaps $\Delta_{uv}(E)$ and $\Delta_{vv}(E)$ are listed in Table \ref{gap_value_SC}.
The values of $\Delta_{uv}(E)$ and $\Delta_{vv}(E)$ of
 the lowest energy resonance $[310]\frac{1}{2}$ is consistent with the above simple
 estimate $\Delta_i^{(eff)} \approx 1.3$ MeV, and we confirm again the
 weak binding does not cause significant change in
 the effective pairing gap of the $[310]\frac{1}{2}$  state.

In order to clarify the behavior of the effective pairing gap of the
weakly bound or resonant $[310]\frac{1}{2}$ state, we compare again 
with the calculations where the pair potential is replaced by
the fixed phenomenological one (cf. Eq.(\ref{Analytic_volume})).
The result is shown in the bottom panel of Fig.~\ref{spectrum_varWS_SCvsanalytic}.
The peak energy of the lowest energy quasiparticle
state $[310]\frac{1}{2}$ varies as  the Woods-Saxon potential depth
$V_{ws}^{0}$ is shifted, and we see that the effective pairing gap 
$\Delta^{(eff)}_{i}$ decreases as the Woods-Saxon potential depth
becomes shallower. The reduction of the pairing gap of the $[310]\frac{1}{2}$ state caused by the
weak binding is seen more explicitly in 
$\Delta_{uv}(E)$ and $\Delta_{vv}(E)$ listed in Table~\ref{gap_value_SC}. 
This reduction of the effective pairing gap due to the weak binding effect
has been pointed out in Refs.\cite{hamamoto_def_HFB_1,hamamoto_def_HFB_2}, where the mechanism of
the reduction is ascribed to a sort of decoupling, i.e, the mechanism 
that the spatially extended wave
function of the weakly bound or resonant quasiparticle state has less overlap
with the pair potential. But our analysis indicates that the
reduction is a consequence of the use of the fixed phenomenological
pair potential. The reduction of the effective
pairing gap does not show up if we describe the pair correlation 
using the selfconsistent pair potential. This is because the weak binding has two
effects, one causing the enhancement of the pair correlation and the
other causing the decoupling of the wave function, which have a tendency to 
compensate each
other.

\section{Conclusions}

We have given a new formulation of the deformed continuum HFB method,
which can be applied to deformed nuclei near the drip-lines. 
The kernel of this formulation is the use of the exact quasiparticle
Green's function constructed on the basis of the coupled-channel representation
in the partial wave expansion. This enables us to impose the correct asymptotics on 
the quasiparticle wave functions of   
the weakly bound and continuum quasiparticle states.
 Consequently we can describe 
the quasiparticle states above the neutron separation energy without
energy discretization. 
 
We have shown several numerical examples to illustrate 
effects of the weak binding and the continuum coupling on
the pair correlation and the quasiparticle spectrum of neutrons 
in deformed nuclei near the neutron drip-line. The calculations are
performed for $^{38}$Mg which is chosen as an example of
prolately deformed nuclei. 
It is found that
there arises
a significant contribution to the pair correlation from the
non-resonant part of the continuum quasiparticle states, and 
the contribution grows as the neutron Fermi energy approaches zero.
This trend is most significant for the quasiparticle states 
with $\Omega=1/2$.
We confirm the strong effects of the continuum coupling
on the $\Omega=1/2$ states\cite{hamamoto_def_HFB_2,
hamamoto_def_WS} causing a steep increase of the width of the resonances, 
but we found that
the contribution to the pair correlation never reduces even with 
the large continuum coupling effects. We also found
that the effective pairing gap of the weakly bound
or resonant $\Omega=1/2$ orbits stay at sizable values.
The effect of the decoupling from 
the pair correlation\cite{hamamoto_def_HFB_1,hamamoto_def_HFB_2} 
is not large, and it may be compensated by the enhancement of the
pair correlation due to the weak binding. The use of the selfconsistent
pair potential is crucial to describe this feature.

\begin{acknowledgments}
The authors thank K. Hagino for useful discussions on the coupled-channel formalism.
This work is supported by
the Grant-in-Aid for Scientific Research (No. 20540259) from the Japan
 Society for the Promotion of Science,  and also by 
the JSPS Core-to-Core Program, International
Research Network for Exotic Femto Systems (EFES). 
\end{acknowledgments}


\end{document}